%% file: Samokhvalov_3_3.tex
\documentclass[11pt,twoside]{gsp}

\input somedef

\begin{document}

\thispagestyle{plain}

\title[About the Symmetry of General Relativity]{About the symmetry of general relativity}
\author[Serhii Samokhvalov]{Serhii Samokhvalov}

\date{}

\maketitle

\comm{Communicated by XXX}\

\begin{abstract}
Generalized deformed gauge groups are used for investigation of symmetry of general relativity (GR). GR is formulated in generalized reference frames, which are represented by affine frames. The general principle of relativity is extended to the requirement of invariance of the theory with respect to the group $GL^g$ of local linear transformations of affine frames. GR is interpreted as the gauge theory of the gauge group of translations $T^g_M$. The groups $GL^g$ and $T^g_M$ are united into the group $S^g_M$, which is their semidirect product and is the complete symmetry group of GR. By $GL^g$-gauge fixing one can obtain: Einstein gravity, GR in an orthogonal frame or teleparallel equivalent of GR, dilaton gravity, unimodular gravity, etc.
\\[0.2cm]
\end{abstract}

\label{first}
\vspace*{1.2cm}
{\renewcommand{\baselinestretch}{0.7}\small\tableofcontents}\normalsize

\section{Introduction}
There are large number theories of gravity locally equivalent to the general relativity. These theories differ in number of field variables and the more or less broad groups of gauge symmetries \cite{Gielen}. A broader group of gauge symmetry gives more opportunities to choose physically acceptable global solutions, for example devoid of singularities.

General relativity falls out of the general scheme of theories of gauge interactions \cite{Blag}. Discussions about the gauge group that should form the basis of the gauge theory of gravity, as well as the geometric structure that should be consistent with it, continue today \cite{Font}, \cite{Pereira}. One might even find an assertion that general relativity is not a gauge theory and does not have a gauge group because any theory can be written in a covariant form, and so diffeomorphisms are empty of dynamical meaning \cite{Andr}.

The main reason for the difficulty of interpreting gravity as a gauge interaction is that the gauge groups of internal symmetry do not act on the space-time manifold, but for gauge gravity this restriction is obviously meaningless because a gravitational field is born by an energy-momentum which associated with space-time translations.

A generalization of gauge groups for the case of their non-trivial action on the space-time continuum was proposed in \cite{Sam1}. Gauge fields, corresponding to an internal or external (space-time) symmetry, both acquire single interpretation as deformation parameters of suitable generalized deformed gauge groups. The most simple and transparent introduction to this subject is given in \cite{SamGG}. The dynamic and geometric meaning of the deformed group of diffeomorphisms with all completeness and mathematical rigor are discussed in \cite{SamR} and \cite{Sam4}.

It should be noted here that group-theoretical description of gauge fields is more fundamental  then geometrical description because according to Klein's Erlangen Program geometrical structures are defined by the groups, which acts on manifolds. Thus, the use of a powerful apparatus of the theory of deformed gauge groups has helped to solve certain long-standing problems that relate to the foundations of geometry \cite{Sam4}.

In this work we use generalized deformed gauge groups for investigation of symmetry of general relativity. By the general relativity (GR) we understand the theory of gravity in the (pseudo)Riemannian space-time with the Hilbert's Lagrangian. However, we shall not restrict ourselves to holonomic reference frames associated with certain coordinate systems as in Einstein gravity (we shall call such coordinates Lagrangian coordinates), and we shall consider a theory of gravity in general anholonomic reference frames (affine frames or affine tetrads).

Transitions between affine frames form the gauge group of linear transformations $G{{L}^{g}}$. The general relativity in an affine frame (GRAF) is invariant under these $G{{L}^{g}}$-transformations. Transitions between holonomic reference frames (or between corresponding Lagrangian coordinates) form the group ${{H}^{g}}$ that is isomorphic to the group of space-time diffeomorphisms and is a special subgroup of the group $G{{L}^{g}}$. In this sense \textit{GRAF is based on the more wide principle of relativity than the Einstein gravity (EG)} \cite{Sam2}.

On the space-time manifold $M$ one can define Eulerian coordinates for numeration of its points. The group $T_{M}^{g}$ of gauge translations on $M$ is the group of diffeomorphisms of Eulerian coordinates in the additive parameterization \cite{SamR}. GRAF is formulated as a gauge theory of translations and so is $T_{M}^{g}$- invariant.

Note that Eulerian coordinates have a different physical meaning unlike Lagrangian ones. If Eulerian coordinates numerate points on the space-time manifold, then Lagrangian ones define the holonomic reference frames. So $T_{M}^{g}$-transformations and ${{H}^{g}}$-transformations have a different physical meaning of gauge translations and changes of holonomic reference frames respectively. The identification of these transformations leads to known difficulties in determining the energy-momentum tensor in EG. On the other hand, separation of gauge translations and changes of reference frames (which takes place in GRAF) permit to define the energy-momentum of gravitational field as a $T_{M}^{g}$-tensor \cite{Moller},  \cite{Duan}. When changing the reference frame (under the $G{{L}^{g}}$-transformations or ${{H}^{g}}$-transformations in the holonomic case), the energy-momentum varies according to the non-tensor law, as it should be \cite{Maluf}. It should be noted that the terms "Lagrangian and Eulerian coordinates" in the space-time continuum are similar (but not identical) to the corresponding concepts in the classical continuum mechanics. The analogy is that at translations the values of Eulerian coordinates change, and the values of Lagrangian coordinates does not.

The gauge group of linear transformations $G{{L}^{g}}$ of affine frames and the generalized gauge group of translations $T_{M}^{g}$  are united to the generalized gauge group $S_{M}^{g}=G{{L}^{g}}\times\!\!\!|\, T_{M}^{g}$ that is their semidirect product. The group $S_{M}^{g}$ is the group of symmetry of GRAF that defines a structure of GRAF.

In this article, first of all, in Section 2, we prove the fact that the currents which are conserved as a result of any generalized gauge symmetries $G_{M}^{g}$ are quasilocal that means they have superpotentials. Moreover, we show that equations for the gauge fields in the gauge theories of generalized gauge groups $G_{M}^{g}$ are equivalent to the expressions of currents (which are conserved as result of $G_{M}^{g}$-symmetries) by the superpotentials.

Then, in Section 3, we describe GRAF: field variables, the Lagrangian, field equations. It is shown, that Palatini equation (equality to zero of variational derivatives of Lagrangian with respect to connection coefficients) is equivalent to metricity and non-torsionity conditions and performed in (pseudo)Riemannian space-time automatically. It is also shown that the equation of gravitational field in GRAF is the Einstein equation in an affine frame.

In Section 4 we study consequences of $S_{M}^{g}$-invariance of GRAF. This is the central part of the work. Here we show that the Palatini equation is a strong Noether identity which follows from the $G{{L}^{g}}$-invariance of GRAF. Then we find a tensor density of energy-momentum of gravitational field in GRAF and its superpotential. Noether identities which follow from the gauge translational invariance ($T_{M}^{g}$-invariance) of GRAF are written down. Equations of gravity written down in the form that is similar to the electromagnetic ones (as Maxwell equations). It is also shown that the complete symmetry group of GRAF $S_{M}^{g}=G{{L}^{g}}\times\!\!\!|\, T_{M}^{g}$ has many deformed forms that lead to a lot of expressions for energy-momentum tensors of gravitational field. Among these forms is one highlighted, so called group of parallel transports in the Riemannian space \cite{Sam4}, that corresponds to the case of reconciling the geometric structure given by the action of the group with the geometric structure of the Riemannian space. The energy-momentum of gravitational field in this case is localizable.

In section 5 we show that by limiting admissible reference frames (by $G{{L}^{g}}$-gauge fixing) from GRAF, in addition to EG, one can obtain other local equivalent formulations of GR: general relativity in an orthonormal frame (GROF) \cite{Duan}, \cite{Rodichev} or teleparallel equivalent of general relativity (TEGR) \cite{Moller}, \cite{Andr}, \cite{Maluf}, dilaton gravity (DG) \cite{Dirac}, unimodular gravity (UG) \cite{Unruh}, etc.

\section{Quasilocality of Gauge Charges}
	In this section it is proved that \textit{quantities which are conserved as a result of gauge symmetries are quasilocal} \cite{Sam3}. Quasilocality of the quantity $Q$ means, that the current ${{J}^{\mu }}$, that relate to the quantity $Q=\int{{{J}^{0}}}dV$, has the form $J_{{}}^{\mu }={{\partial }_{\sigma }}S_{{}}^{\sigma \mu }$, where $S_{{}}^{\sigma \mu }=-S_{{}}^{\mu \sigma }$ is its superpotential \cite{Szabados}. Here ${{x}^{\mu }}$ are the Eulerian coordinates on the space-time manifold $M$, for which we will use the Greek indices of the middle of the alphabet. Our proof is based on the generalized Noether theorem, which generalizes the former Noether theorem \cite{Noether} for the case of existence of a surface part of the Lagrangian.

	Let the system of fields ${{q}^{i}}(x)$ (with indices of the middle of the Latin alphabet) be given on the space-time manifold $M$. Let the action be given: $S=\int{\Lambda \,dx}$, where $dx$ is an element of coordinate volume in $M$, with the Lagrangian
\begin{equation}\label{eq1}
\Lambda =L+{{\partial }_{\sigma }}{{V}^{\sigma }}
\end{equation}
which consists of the volume $L$ and the surface ${{\partial }_{\sigma }}{{V}^{\sigma }}$ parts (${{\partial }_{\sigma }}:=\partial /\partial {{x}^{\sigma }}$), and $L$ and ${{V}^{\sigma }}$ depend on the fields $q$ and their partial derivatives $\partial q$. Therefore, the complete Lagrangian is linearly dependent also on the second derivatives of the fields $\partial {{}^{2}}q$, which fold in ${{\partial }_{\sigma }}{{V}^{\sigma }}$. The surface part ${{\partial }_{\sigma }}{{V}^{\sigma }}$ does not affect the equations of motion determined solely by the bulk Lagrangian $L$. The surface part characterizes the boundary-value problem as well as the symmetry defined by the boundary-value problem.

	At the transformations
$$
\begin{array}{rcl}
	{{x}'}^{\mu } = {{x}^{\mu }}+\delta {{x}^{\mu}},\qquad
      q{{'}^{i}}(x) = {{q}^{i}}(x)+\delta {{q}^{i}}
\end{array}
$$
the action is transformed: $\bar{\delta }S=\int{{\delta }'\Lambda dx}$, where ${\delta }'\Lambda :={\Lambda }'({x}')J-\Lambda (x)$ is the integral variation of the Lagrangian $\Lambda $, $J:=\left| {{\partial }_{\rho }}{{{{x}'}}^{\sigma }} \right|$. Here ${\Lambda }'({x}')$ is the Lagrangian, calculated with $q{{'}^{i}}({x}')$. According to (\ref{eq1}) we obtain
\begin{equation}\label{eq2}
{\delta }'\Lambda ={\delta }'L+{{\partial }_{\sigma }}{\delta }'{{V}^{\sigma }}
\end{equation}
where
\begin{equation}\label{eq3}
{\delta }'{{V}^{\sigma }}:={{\partial }_{{{\rho }'}}}{{x}^{\sigma }}{{{V}'}^{{{\rho }'}}}({x}')J-{{V}^{\sigma }}(x).
\end{equation}
Here we have taken into account that ${{\partial }_{\sigma }}({{\partial }_{{{\rho }'}}}{{x}^{\sigma }}J)\equiv 0$, where ${{\partial }_{{{\rho }'}}}:=\partial /\partial {{{x}'}^{{{\rho }'}}}$. We obtain
$$
\begin{array}{rcl}
	{\delta }'L={{\nabla }_{{{q}^{i}}}}L\delta {{q}^{i}}+{{\partial }_{\sigma }}(P_{i}^{\sigma }\delta {{q}^{i}}+L\delta {{x}^{\sigma }})
\end{array}
$$
where
\begin{equation}\label{eq4}
{{\nabla }_{{{q}^{i}}}}L:={{Q}_{i}}-{{\partial }_{\sigma }}P_{i}^{\sigma }
\end{equation}
is the variational derivatives of the volume Lagrangian, ${{Q}_{i}}:={{\partial }_{{{q}^{i}}}}L$, $P_{i}^{\sigma }:={{\partial }_{{{\partial }_{\sigma }}{{q}^{i}}}}L$, where ${{\partial }_{{{q}^{i}}}}:=\partial /\partial {{q}^{i}}$, ${{\partial }_{{{\partial }_{\nu }}{{q}^{i}}}}:=\partial /\partial {{\partial }_{\nu }}{{q}^{i}}$. Thus, taking into account (\ref{eq2}), we obtain
\begin{equation}\label{eq5}
{\delta }'\Lambda ={{\nabla }_{{{q}^{i}}}}L\delta {{q}^{i}}+{{\partial }_{\sigma }}(P_{i}^{\sigma }\delta {{q}^{i}}+L\delta {{x}^{\sigma }}+{\delta }'{{V}^{\sigma }}).
\end{equation}

	Let be given generalized gauge group $G_{M}^{g}$  \cite{Sam1}, that is parameterized by the functions ${{g}^{a}}(x)$ (here we will use indexes from the beginning of the Latin alphabet) and its infinitesimal transformations ${{\delta }_{g}}$ are defined	
\begin{equation}\label{eq8}
{{\delta }_{g}}{{x}^{\mu }}=h_{a}^{\mu }{{g}^{a}},\qquad
{{\delta }_{g}}{{q}^{i}}=a_{a}^{i}{{g}^{a}}+b_{\,\,a}^{i\mu }{{\partial }_{\mu }}{{g}^{a}},\qquad
{{{\delta }'}_{g}}{{V}^{\sigma }}=c_{a}^{\sigma }{{g}^{a}}+d_{a}^{\mu \sigma }{{\partial }_{\mu }}{{g}^{a}}
\end{equation}
where $a_{a}^{i}$, $b_{\,\,a}^{i\mu }$, $c_{a}^{\sigma }$, $d_{a}^{\mu \sigma }$ and $h_{a}^{\mu }$ are the functions, that depend on $x$ explicitly or implicitly via the fields $q$ and their derivatives. Formulas (\ref{eq8}) define the group $G_{M}^{g}$ as the transformations group. We will omit the display of dependence on $x$ of the fields $q^i$ and the parameters $g^a$ of the group $G_{M}^{g}$, where this does not lead to confusion.

	Substitution of expressions (\ref{eq8}) in (\ref{eq5}) allows us to express the integral variation of the total Lagrangian through the parameters of the group $G_{M}^{g}$ and their derivatives
\begin{equation}\label{eq9}
{{{\delta }'}_{g}}\Lambda ={{\nabla }_{{{q}^{i}}}}L\,(a_{a}^{i}{{g}^{a}}+b_{\,\,a}^{i\mu }{{\partial }_{\mu }}{{g}^{a}})-{{\partial }_{\sigma }}(J_{a}^{\sigma }{{g}^{a}}+S_{a}^{\mu \sigma }{{\partial }_{\mu }}{{g}^{a}})
\end{equation}
where the definition is entered
\begin{equation}\label{eq10}
J_{a}^{\sigma }:=-P_{i}^{\sigma }a_{a}^{i}-Lh_{a}^{\sigma }-c_{a}^{\sigma },\qquad
S_{a}^{\mu \sigma }:=-P_{i}^{\sigma }b_{\,\,a}^{i\mu }-d_{a}^{\mu \sigma }
\end{equation}
for the generalized Noether current $J_{a}^{\sigma }$ and its superpotential $S_{a}^{\mu \sigma }$.

	In the case of invariance of the action $S$ with respect to the transformations of the group $G_{M}^{g}$, we have ${{\bar{\delta }}_{g}}S=0$ and hence ${{{\delta }'}_{g}}\Lambda =0$. Since the parameters of the group $G_{M}^{g}$ and their derivatives are independent functions, equating of coefficients at them in expression (\ref{eq9}) to zero allows us to find the following generalized strong (performed everywhere, not only on shell) Noether identities
\begin{equation}\label{eq12}
{{\nabla }_{{{q}^{i}}}}L\,a_{a}^{i}={{\partial }_{\sigma }}J_{a}^{\sigma },\qquad
{{\nabla }_{{{q}^{i}}}}L\,b_{\,\,a}^{i\mu }=J_{a}^{\mu }+{{\partial }_{\sigma }}S_{a}^{\mu \sigma },\qquad
S_{a}^{\mu \nu }=-S_{a}^{\nu \mu }.
\end{equation}

Using expression (\ref{eq4}) for the variational derivatives, second identity (\ref{eq12}) can be represented as
\begin{equation}\label{eq15}
J_{a}^{\mu }={{Q}_{i}}b_{\,\,a}^{i\mu }+P_{i}^{\sigma }{{\partial }_{\sigma }}b_{\,\,a}^{i\mu }+{{\partial }_{\sigma }}d_{\,\,\,a}^{\mu \sigma }
\end{equation}
which often makes it easier to find the current $J_{a}^{\mu }$, especially in the case of $b_{\,\,a}^{i\mu }=const$ and $d_{\,\,\,a}^{\mu \sigma }=const$, thus $J_{a}^{\mu }={{Q}_{i}}b_{\,\,a}^{i\mu }$.

On shell second identity (\ref{eq12}) gives
$$
\begin{array}{rcl}
	J_{a} ^{\mu}=- \partial _\sigma S_{a}^{\mu \sigma}
\end{array}
$$
so \textit{current} $J_{a}^{\mu }$ \textit{is quasilocal and} $S_{a}^{\mu \sigma }$ \textit{is its superpotential}.

In gauge theories part of fields ${{q}^{i}}$ are the gauge fields $A_{\mu }^{a}$, for them $b_{a\mu }^{\nu b}\sim\delta _{a}^{b}\delta _{\mu }^{\nu }$ and others coefficients $b_{\,\,a}^{i\mu }$ equal to zero. In this case $J_{a}^{\mu }\sim{{\partial }_{A_{\mu }^{a}}}L$. Moreover, second identity (\ref{eq12}) gives
$$
\begin{array}{rcl}
	{{\nabla }_{A_{\mu }^{a}}}L\sim J_{a} ^{\mu}+ \partial _\sigma S_{a}^{\mu \sigma}
\end{array}
$$
so  \textit{equation of motion for gauge fields} ${{\nabla }_{A_{\mu }^{a}}}L=0$  \textit{is reduced to an expression of the current by the superpotential}: $J_{a}^{\mu }=-{{\partial }_{\sigma }}S_{a}^{\mu \sigma }$. From this formula, taking into account the antisymmetry of the superpotential by the upper indices (\ref{eq12}), immediately follows the weak Noether identity ${{\partial }_{\sigma }}J_{a}^{\sigma }=0$. That is the conservation law of the current that follows also from first identity (\ref{eq12}) on shell.
	
On the infinitesimal level deformations of the group $G_{M}^{g}\to G_{M}^{gH}$ are reduced to gauge linear transformations of parameters ${{g}^{a}}=H_{\alpha }^{a}{{g}^{\alpha }}$ with $\left| H_{\alpha }^{a} \right|\ne 0$ \cite{Sam1}.
Under deformations currents and superpotentials change according to the formulas
\begin{equation}\label{eq16}
J_{\alpha }^{\mu }=H_{\alpha }^{a}J_{a}^{\mu }+{{\partial }_{\sigma }}H_{\alpha }^{a}S_{a}^{\sigma \mu },\qquad
S_{\alpha }^{\mu \nu }=H_{\alpha }^{a}S_{a}^{\mu \nu }.
\end{equation}

Thus deformations lead to redistributions between $J_{a}^{\mu }$ and ${{\partial }_{\sigma }}S_{a}^{\mu \sigma }$ in the expression $J_{a}^{\mu }+{{\partial }_{\sigma }}S_{a}^{\mu \sigma }$.

\section{General Relativity in an Affine Frame}

In this section we \textit{generalize the general principle of relativity to use anholonomic reference frames (affine frames)}. Here we describe general relativity in an affine frame (GRAF): main field variables of the theory, Lagrangian and field equations \cite{Sam2}.

\subsection{Field variables and Lagrangian of GRAF}

Suppose that in the space-time $M$ an affine frame is given ${{h}_{m}}=h_{m}^{\mu }{{\partial }_{\mu }}$, which vectors are numbered in Latin letters from the middle of the alphabet. Suppose also that space-time is (pseudo) Riemannian, and therefore a scalar product is given ${{h}_{m}}\cdot {{h}_{n}}={{g}_{mn}}$. Replacement of coordinate indices to frame indices and vice versa will be performed using matrices $h_{m}^{\mu }$ and inverse ones $h_{\mu }^{m}$, in particular ${{g}_{\mu \nu }}=h_{\mu }^{m}h_{\nu }^{n}{{g}_{mn}}$. We will also assume that in $M$ is given an affine connection with connection 1-form $\gamma _{\,\,n}^{m}=\gamma _{\,\,\mu n}^{m}\,d{{x}^{\mu }}$, where $\gamma _{\,\,\mu n}^{m}$ is the connection coefficients in the affine frame ${{h}_{m}}$. Anholonomic coefficients $F_{\,\,mn}^{k}\,$ of the reference fields ${{h}_{m}}$ are defined by the expression
$$
\begin{array}{rcl}
[{{h}_{m}},{{h}_{n}}]=F_{\,\,mn}^{k}\,{{h}_{k}}
\end{array}
$$
where square brackets denote the commutator of the vector fields. The group-theoretic meaning of the affine frame as a set of generators of a deformed group of diffeomorphisms, and of the anholonomic coefficients as its structural functions, similar to the structural constants of the Lie group, is important \cite{Sam1}.

	The condition of absence of torsion (non-torsionity condition) is
\begin{equation}\label{eq18}
F_{\,\,mn}^{k}=\gamma _{\,\,mn}^{k}-\gamma _{\,\,nm}^{k}
\end{equation}
and the metricity condition is
\begin{equation}\label{eq19}
{{\partial }_{k}}{{g}_{mn}}=\gamma _{mkn}^{\cdot }+\gamma _{nkm}^{\cdot }
\end{equation}
where ${{\partial }_{k}}{{g}_{mn}}:=h_{k}^{\mu }{{\partial }_{\mu }}{{g}_{mn}}$. Under the conditions (\ref{eq18}), (\ref{eq19}), the connection coefficients are decomposed into the sum $\gamma _{mkn}^{\cdot }=\omega _{mkn}^{\cdot }+\sigma _{mkn}^{\cdot }$, where
\begin{equation}\label{eq20}
\omega _{mkn}^{\cdot }=\tfrac{1}{2}(F_{k\,mn}^{\cdot }+F_{mkn}^{\cdot }-F_{nkm}^{\cdot }),\quad
\sigma _{mkn}^{\cdot }=\tfrac{1}{2}({{\partial }_{n}}{{g}_{mk}}+{{\partial }_{k}}{{g}_{mn}}-{{\partial }_{m}}{{g}_{kn}})
\end{equation}
with symmetry properties $\omega _{mkn}^{\cdot }=-\omega _{nkm}^{\cdot }$, $\sigma _{mnk}^{\cdot }=\sigma _{mkn}^{\cdot }$. The point indicates from where the index was lowered (or upped).

The main field variables in GRAF are the transition coefficients $h_{\mu }^{m}$ between affine and coordinate frames, the metric ${{g}_{mn}}$ and the connection coefficients $\gamma _{\,\,\mu n}^{m}$ in the affine frame (frame-metric-affine gravity theory: generalized Einstein-Palatini model). This choice of field variables dictates the breakup of the Hilbert's Lagrangian
\begin{equation}\label{eq22}
\Lambda =-\tfrac{1}{2\kappa }eR
\end{equation}
to a bulk part ${{L}_{\gamma }}$  (the truncated Hilbert's Lagrangian) that does not depend on the derivatives of the connection coefficients $\gamma _{\,\,\mu n}^{m}$, and a surface part having the form of divergence ${{\partial }_{\sigma }}V_{\gamma }^{\sigma }$ into which derivatives of $\gamma _{\,\,\mu n}^{m}$ are folded
\begin{equation}\label{eq23}
\Lambda ={{L}_{\gamma }}+{{\partial }_{\sigma }}V_{\gamma }^{\sigma }
\end{equation}
where $\kappa $ is the Einstein gravitational constant (which we will hereafter assume equal to 1), $e:=\sqrt{|{{g}_{\mu \nu }}|}$, and $R$ is the scalar curvature of the space-time $M$. Indeed, since
\begin{equation}\label{eq24}
\tfrac{1}{2}eR=\Sigma _{m}^{\ \ n\mu \nu }({{\partial }_{\mu }}\,\gamma _{\nu \,n}^{m}+\gamma _{\mu \,s}^{m}\,\gamma _{\nu \,n}^{s})
\end{equation}
where $\Sigma _{m}^{\ \ n\mu \nu }:=\tfrac{1}{2}e\delta _{ms}^{\mu \nu }{{g}^{sn}}$, $\delta _{mn}^{\mu \nu }:=h_{m}^{\mu }\,h_{n}^{\nu }-h_{n}^{\mu }\,h_{m}^{\nu }$, throwing in the first term of formula (\ref{eq24}) the derivative to the first factor and changing the sign, we obtain
$$
\begin{array}{rcl}
\Lambda ={{\partial }_{\mu }}\Sigma _{m}^{\ \ n\mu \nu }\,\gamma _{\nu \,n}^{m}-\Sigma _{m}^{\ \ n\mu \nu }\gamma _{\mu \,s}^{m}\,\gamma _{\nu \,n}^{s}\ -{{\partial }_{\mu }}(\Sigma _{m}^{\ \ n\mu \nu }\,\gamma _{\nu \,n}^{m}).
\end{array}
$$
So
\begin{align} \label{eq26}
{{L}_{\gamma }}&= {{\partial }_{\mu }}\Sigma _{m}^{\ \ n\mu \nu }\,\gamma _{\nu \,n}^{m}-H_\gamma, \qquad
H_\gamma := \Sigma _{m}^{\ \ n\mu \nu }\gamma _{\mu \,s}^{m}\,\gamma _{\nu \,n}^{s}
\notag\\[-2mm] &\\[-2mm]
V_{\gamma }^{\sigma }&=\Sigma _{m}^{\ \ n\nu \sigma }\,\gamma _{\nu \,n}^{m}=e\,\gamma _{\,n\,\ \cdot }^{[n\,\sigma ]}.
\notag
\end{align}
Square brackets here mean antisymmetrization by the indexes contained therein.

	All addends of the partition (\ref{eq23}), like Hilbert's own Lagrangian (\ref{eq22}), are scalar densities with respect to changing of coordinates ${{x}^{\mu }}$. A general-covariant method of separating a surface term from the Hilbert's Lagrangian is possible due to the use of frames fields that do not change when changing coordinates.

\subsection{Palatini equation}

In this subsection we show that the \textit{Palatini equation obtained by variation of Hilbert's Lagrangian with respect to coefficients of affine connection in the affine frame is the identity due to the metricity and non-torsionity conditions.}

The integral variation of Lagrangian ${{L}_{\gamma }}$ is determined by the expression
\begin{align} \label{eq27}
{\delta }'{{L}_{\gamma }}&={{\partial }_{\gamma _{\ \ \nu n}^{m}}}{{L}_{\gamma }}\,\delta \gamma _{\ \ \nu n}^{m}+{{\nabla }_{h_{\nu }^{n}}}{{L}_{\gamma }}\,\delta h_{\nu }^{n}+{{\nabla }_{{{g}_{mn}}}}{{L}_{\gamma }}\,\delta {{g}_{mn}}
\notag\\[-2mm] &\\[-2mm]
&+{{\partial }_{\mu }}(\delta \Sigma _{m}^{\ \ n\mu \nu }\,\gamma _{\nu \,n}^{m}+{{L}_{\gamma }}\delta {{x}^{\mu }}).
\notag
\end{align}

Let us Consider the terms in formula (\ref{eq27}). The derivatives of the Lagrangian ${{L}_{\gamma }}$ with respect to $\gamma _{\ \ \nu n}^{m}$ are	
$$
\begin{array}{rcl}
{{\partial }_{\gamma _{\ \ \nu n}^{m}}}{{L}_{\gamma }}={{\partial }_{\sigma }}\Sigma _{m}^{\ \ n\sigma \nu }-M_{m}^{\ \ n\nu }
\end{array}
$$
where
$$
\begin{array}{rcl}
M_{m}^{\ \ n\nu }:={{\partial }_{\gamma _{\ \ \nu n}^{m}}}H_{\gamma }=\Sigma _{m}^{\ \ s\nu \sigma }\gamma _{\ \,\sigma \,s}^{n}-\Sigma _{s}^{\ \ n\nu \sigma }\gamma _{\ \,\,\sigma \,m}^{s}.
\end{array}
$$
Equating these derivatives to zero, we obtain equation which we will call the Palatini equation in the affine frame
\begin{equation}\label{eq28}
{{\partial }_{\sigma }}\Sigma _{m}^{\ \ n\sigma \nu }-M_{m}^{\ \ n\nu }=0.
\end{equation}

\textit{With metricity condition} (\ref{eq19}), \textit{Palatini equation} (\ref{eq28}) \textit{goes into the non-torsionity condition} (\ref{eq18}) \textit{and vice versa. } Therefore, when these two conditions are fulfilled simultaneously, Palatini equation is fulfilled as identity.

Indeed, a direct calculation gives
\begin{align} \label{eq29}
{{\partial }_{\sigma }}\Sigma _{m}^{\ \ n\sigma \nu }&=\tfrac{1}{2}e\,[F_{m\,s}^{\nu \,}-\delta _{ms}^{\nu \sigma }({{R}_{\sigma }}+{{S}_{\sigma }})+\delta _{mk}^{\nu \sigma }{{g}^{kp}}{{\partial }_{\sigma }}{{g}_{ps}}]\,{{g}^{sn}}
\notag\\[-2mm] &\\[-2mm]
M_{m}^{\ \ n\nu }&=-\tfrac{1}{2}e\,(\gamma _{s\,m\,\,\cdot }^{\,\cdot \ \ \,\,\nu }+\gamma _{\ \,s\,m}^{\nu }-h_{m}^{\nu }\,\gamma _{sp\,\cdot }^{\,\cdot \ \,\,p}-h_{s}^{\nu }{{\tilde{R}}_{m}}){{g}^{sn}}
\notag
\end{align}
where ${{R}_{m}}:=F_{s\,m}^{s}$, ${{S}_{m}}:=\tfrac{1}{2}{{g}^{ps}}{{\partial }_{m}}{{g}_{ps}}$, ${{\tilde{R}}_{m}}:=\gamma _{\ \,s\,m}^{s}$, so the Palatini equation (\ref{eq28}) is reduced to the equation
\begin{align} \label{eq30}
F_{m\,n}^{\nu \,}+\gamma _{n\,m\,\cdot }^{\,\cdot \ \ \,\,\nu }+\gamma _{\ \,n\,m}^{\nu }&-h_{s}^{\nu }{{g}^{sp}}{{\partial }_{m}}{{g}_{pn}}+h_{n}^{\nu }({{R}_{m}}+{{S}_{m}}-{{\tilde{R}}_{m}})
\notag\\[-2mm] &\\[-2mm]
&-h_{m}^{\nu }({{R}_{n}}+{{S}_{n}}-{{g}^{sp}}{{\partial }_{s}}{{g}_{pn}}+\gamma _{np\,\cdot }^{\,\cdot \ \,\,p})=0.
\notag
\end{align}

Due to the metricity condition (\ref{eq19}), we have ${{\tilde{R}}_{n}}={{g}^{sp}}{{\partial }_{s}}{{g}_{pn}}-\gamma _{np\,\cdot }^{\,\cdot \ \,\,p}$, ${{\tilde{S}}_{m}}={{S}_{m}}$, where ${{\tilde{S}}_{m}}:=\gamma _{\ \,m\,s}^{s}$, and equation (\ref{eq30}) is reduced to the equation
$$
\begin{array}{rcl}
F_{m\,n}^{\nu \,}=\gamma _{\ \,m\,n}^{\nu }-\gamma _{\ \,n\,m}^{\nu }+\delta _{mn}^{\nu \,s}({{R}_{s}}+{{S}_{s}}-{{\tilde{R}}_{s}}).
\end{array}
$$
Summing up this equation by $\nu $ and $m$ (with help of $h_{\nu }^{m}$) we obtain $(d-2)({{R}_{n}}+{{S}_{n}}-{{\tilde{R}}_{n}})=0$, where $d$ is the dimension of space-time $M$. Therefore, when $d\ne 2$ equation (\ref{eq30}) is reduced to the non-torsionity condition (\ref{eq18}).

	Now let the initial assumption be the non-torsionity condition (\ref{eq18}), from which follows ${{R}_{s}}={{\tilde{R}}_{s}}-{{\tilde{S}}_{s}}$. In this case, equation (\ref{eq30}) is reduced to the equation
\begin{eqnarray*}
{{\partial }_{m}}{{g}_{kn}}=\gamma _{km\,n}^{\,\cdot }+\gamma _{n\,m\,k}^{\,\cdot \,\,}&+&{{g}_{kn}}({{S}_{m}}-{{\tilde{S}}_{m}})\\
&-&{{g}_{km}}({{\tilde{R}}_{n}}+{{S}_{n}}-{{\tilde{S}}_{n}}-{{g}^{sp}}{{\partial }_{s}}{{g}_{pn}}+\gamma _{np\,\cdot }^{\,\cdot \,\,p}).
\end{eqnarray*}
Summing up this equation by $k$ and $m$ (with help of ${{g}^{km}}$) we obtain
$$
\begin{array}{rcl}
	(1-d)\,({{\tilde{R}}_{n}}+{{S}_{n}}-{{\tilde{S}}_{n}}-{{g}^{sp}}{{\partial }_{s}}{{g}_{pn}}+\gamma _{np\,\cdot }^{\,\cdot \ \,\,p})=0
\end{array}
$$
so (for $d\ne 1$) previous equation is reduced to the equation
$$
\begin{array}{rcl}
	{{\partial }_{m}}{{g}_{kn}}=\gamma _{km\,n}^{\,\cdot }+\gamma _{n\,m\,k}^{\,\cdot \ \ \,\,}+{{g}_{kn}}({{S}_{m}}-{{\tilde{S}}_{m}}).
\end{array}
$$
And at last, summing up this equation by $k$ and $n$ (with help of ${{g}^{kn}}$) we obtain $(d-2)({{S}_{m}}-{{\tilde{S}}_{m}})=0$, so for $d>2$ this equation is reduced to the metricity condition (\ref{eq19}).

Thus, due to the Palatini equation in the affine frame (\ref{eq28}), the metricity condition (\ref{eq19}) and the non-torsionity condition (\ref{eq18}) follow one from the other.

In what follows, we assume fulfilling conditions (\ref{eq18}) and (\ref{eq19}) (which always take place in the Riemannian space), and hence the Palatini equation (\ref{eq28}).

\subsection{Field equations}

In this subsection we show that the \textit{field equation in GRAF is the Einstein equation in an affine frame. }

Both variational derivatives in formula (\ref{eq27}) are expressed by the symmetric Einstein tensor in our case (under the conditions (\ref{eq18}) and (\ref{eq19})) ${{G}_{mn}}={{R}_{mn}}-\tfrac{1}{2}{{g}_{mn}}R$
$$
\begin{array}{rcl}
	{{\nabla }_{h_{\nu }^{n}}}{{L}_{\gamma }}= eG_{n}^{\nu }, \qquad
        {{\nabla }_{{{g}_{mn}}}}{{L}_{\gamma }}= \tfrac{1}{2}e{{G}^{mn}}
\end{array}
$$
so $h_{\nu }^{m}{{\nabla }_{h_{\nu }^{n}}}{{L}_{\gamma }}=2{{g}_{ns}}{{\nabla }_{{{g}_{ms}}}}{{L}_{\gamma }}$. This indicates that the number of fields $h_{\mu }^{m}$ and ${{g}_{mn}}$ is redundant to find the equation of motion of the gravitational field in GRAF. For example, we can choose $h_{\mu }^{m}$ as basic field variables and consider ${{g}_{mn}}$ as parameters. In particular, ${{g}_{mn}}$ may be a metric ${{\eta }_{mn}}$ of a flat space-time corresponding to the choice of orthonormal frame fields, or a metric of any other (pseudo)Riemannian space-time, such as AdS space-time. The space-time with the metric ${{g}_{mn}}$ can be considered as a vacuum, in which gravitational phenomena are described by the potentials $h_{\mu }^{m}$. In the group-theoretical interpretation of gravitational field as a gauge field \cite{Sam1}, the gravitational field acts as a further deformation of the background (vacuum) space-time.

	All addends with variational derivatives in formula (\ref{eq27}) are reduced to one
\begin{equation}\label{eq31}
eG_{n}^{\nu }\delta h_{\nu }^{n}+\tfrac{1}{2}e{{G}^{mn}}\,\delta {{g}_{mn}}=\tfrac{1}{2}e{{G}^{\mu \nu }}\,\delta {{g}_{\mu \nu }}.
\end{equation}

Since $\Sigma _{m}^{\ \ n\mu \nu }$ depends only on the fields $h_{\sigma }^{s}$ and ${{g}_{ps}}$ and not on their derivatives, we have
$$
\begin{array}{rcl}
\delta \Sigma _{m}^{\ \ n\mu \nu }={{\partial }_{h_{\sigma }^{s}}}\Sigma _{m}^{\ \ n\mu \nu }\delta h_{\sigma }^{s}+{{\partial }_{{{g}_{ps}}}}\Sigma _{m}^{\ \ n\mu \nu }\delta {{g}_{ps}}.
\end{array}
$$
Directly from the definition of $\Sigma _{m}^{\ \ n\mu \nu }$ we find
\begin{align} \label{eq32}
{{\partial }_{h_{\sigma }^{s}}}\Sigma _{m}^{\ \ n\mu \nu }&=h_{s}^{\sigma }\Sigma _{m}^{\ \ n\mu \nu }+h_{s}^{\mu }\Sigma _{m}^{\ \ n\nu \sigma }+h_{s}^{\nu }\Sigma _{m}^{\ \ n\sigma \mu }
\notag\\[-2mm] &\\[-2mm]
{{\partial }_{{{g}_{ps}}}}\Sigma _{m}^{\ \ n\mu \nu }&=-{{g}^{n\{p}}\Sigma _{m}^{\ \ s\}\mu \nu }+\tfrac{1}{2}{{g}^{ps}}\Sigma _{m}^{\ \ n\mu \nu }.
\notag
\end{align}
Braces here mean symmetrization by the indexes contained therein. So
$$
\begin{array}{rcl}
\delta \Sigma _{m}^{\ \ n\mu \nu }\,\gamma _{\nu \,n}^{m}=B_{s}^{\sigma \mu }\delta h_{\sigma }^{s}+\tfrac{1}{2}D_{{}}^{ps\,\mu }\delta {{g}_{ps}},
\end{array}
$$
where
\begin{align} \label{eq34}
B_{s}^{\sigma \mu }:&={{\partial }_{{{\partial }_{\mu }}h_{\sigma }^{s}}}{{L}_{\gamma }}={{\partial }_{h_{\sigma }^{s}}}\Sigma _{m}^{\ \ n\mu \nu }\gamma _{\nu \,n}^{m}
\notag\\[-2mm] &\\[-2mm]
\tfrac{1}{2}D_{{}}^{ps\,\mu }:&={{\partial }_{{{\partial }_{\mu }}{{g}_{ps}}}}{{L}_{\gamma }}={{\partial }_{{{g}_{ps}}}}\Sigma _{m}^{\ \ n\mu \nu }\gamma _{\nu \,n}^{m}
\notag
\end{align}
- generalized momentums of fields $h_{\sigma }^{s}$ and ${{g}_{ps}}$ respectively. Using (\ref{eq32}) we obtain
\begin{align} \label{eq36}
B_{s}^{\sigma \mu }&=e\gamma _{\ \ \,s\ .}^{[\sigma \ \mu ]}-h_{s}^{\sigma }\,V_{\gamma }^{\mu }+h_{s}^{\mu }\,V_{\gamma }^{\sigma }
\notag\\[-2mm] &\\[-2mm]
D_{{}}^{ps\,\mu }&=-e\gamma _{\ \quad \!\! \cdot \ \! \,\cdot }^{\mu \{ps\}}+e\,\gamma _{n\ \ \,\cdot }^{n\,\{p}{{g}^{s\}\mu }}-{{g}^{ps}}V_{\gamma }^{\mu }.
\notag
\end{align}

Formula (\ref{eq27}), taking into account (\ref{eq28}), (\ref{eq31}) and (\ref{eq34}), takes the form
\begin{equation}\label{eq38}
{\delta }'{{L}_{\gamma }}=\tfrac{1}{2}\,e{{G}^{\mu \nu }}\delta {{g}_{\mu \nu }}+{{\partial }_{\mu }}(B_{s}^{\sigma \mu }\delta h_{\sigma }^{s}+\tfrac{1}{2}D_{{}}^{ps\,\mu }\delta {{g}_{ps}}+{{L}_{\gamma }}\delta {{x}^{\mu }}).
\end{equation}
It is interesting to note that generalized momentums can be found solely from the vector density $V_{\gamma }^{\mu }$, which determines the surface part of Hilbert's Lagrangian
$$
\begin{array}{rcl}B_{s}^{\sigma \mu }=-{{\partial }_{h_{\sigma }^{s}}}V_{\gamma }^{\mu }, \qquad   \tfrac{1}{2}D_{{}}^{ps\,\mu }=-{{\partial }_{{{g}_{ps}}}}V_{\gamma }^{\mu }.
\end{array}
$$

We will note the formula
$$
\begin{array}{rcl}
M_{m}^{\ \ n\sigma }=D_{m}^{\,\cdot \,\,n\,\sigma }-B_{m}^{n\sigma }
\end{array}
$$
which is proved with the help of formulas (\ref{eq29}), (\ref{eq36}) by the direct substitution.

From the Palatine equation (\ref{eq28}) follows $M_{m}^{\ \ n\nu }\gamma _{\nu \,n}^{m}={{\partial }_{\mu }}\Sigma _{m}^{\ \ n\mu \nu }\,\gamma _{\nu \,n}^{m}=2H_\gamma$, so, provided that it is satisfied
\begin{equation}\label{eq39}
{{L}_{\gamma }}=H_\gamma=\Sigma _{m}^{\ \ n\mu \nu }\gamma _{\mu \,s}^{m}\,\gamma _{\nu \,n}^{s}.
\end{equation}
In a coordinate frame, when $\gamma _{\mu \,n}^{s}$ is converted to Christoffel symbols $\Gamma _{\mu \,\nu }^{\sigma }$, the Lagrangian ${{L}_{\gamma }}$ coincides with the truncated Hilbert's Lagrangian ${{L}_{\Gamma }}$ \cite{Landau}. In an orthonormal frame, when $\gamma _{\mu \,n}^{m}$ goes into the Ricci rotational coefficients $\Omega _{\mu n}^{m}$, the Lagrangian ${{L}_{\gamma }}$ is the same as Lagrangian ${{L}_{\Omega }}$ that was suggested in \cite{Duan}, \cite{Cho} (this Lagrangian is used in the teleparallel equivalent of general relativity (TEGR), too  \cite{Maluf}). In the general case of an affine frame, when applying formula (\ref{eq39}) to the Lagrangian  ${{L}_{\gamma }}$, conditions (\ref{eq18}) and (\ref{eq19}) must be taken into account. Therefore, the connection coefficients will no longer be arbitrary in this case, and formulas (\ref{eq20}) for their constituents must be applied when they are varied.

\section{Consequences of Gauge Invariance of GRAF}
Hilbert's Lagrangian $\Lambda $ exhibits uniquely wide gauge symmetry. First of all, $\Lambda $ is invariant with respect to $G{{L}^{g}}$-transformations of the affine frame fields, which corresponds to the transition between general reference frames. In addition, $\Lambda $ is invariant with respect to the gauge translations of the space-time, that is, $T_{M}^{g}$-transformations. Both these groups combine into a semidirect product $S_{M}^{g}=G{{L}^{g}}\times\!\!\!|\, T_{M}^{g}$, forming a complete symmetry group of GRAF. This symmetry of GRAF is a consequence of the fact that when transforming frame fields
\begin{equation}\label{eq40}
{{{h}'}_{{{m}'}}}({x}')=L_{{{m}'}}^{\ \ n}(x)\,{{h}_{n}}(x),\qquad
{{{x}'}^{\mu }}={{x}^{\mu }}+{{t}^{\mu }}(x)
\end{equation}
which are parameterized by the functions ${{g}^{a}}(x)=(L_{{{m}'}}^{\ \ n}(x),{{t}^{\mu }}(x))\in S_{M}^{g}$, the scalar curvature remains unchanged ${R}'({x}')=R(x)$, and hence ${{{\delta }'}_{g}}\Lambda ={\Lambda }'({x}')J-\Lambda (x)=0$. As a result, the equations, which are obtained from the Lagrangian${{L}_{\gamma }}$, are $S_{M}^{g}$-invariant.
	
Transformations (\ref{eq40}) uniquely determine the way in which the semidirect product of the groups $G{{L}^{g}}$ and $T_{M}^{g}$ is formed, defining action $\phi_{{{t}^{\mu }}(x)} L_{{{m}'}}^{\ \ n}(x)=L_{{{m}'}}^{\ \ n}(x+t(x))$. The transformations from the group $S_{M}^{g}$ for the field variables of GRAF are
\begin{align} \label{eq42}
{{h}'}_{\mu }^{{{m}'}}({x}')&=L_{\ \,k}^{{{m}'}}(x)\,h_{\nu }^{k}(x){{{\partial }'}_{\mu }}{{x}^{\nu }},\quad
{{{g}'}_{{m}'{n}'}}({x}')=L_{{{m}'}}^{\ \ k}(x)L_{{{n}'}}^{\ \ l}(x){{g}_{kl}}(x)
\notag\\[-2mm] &\\[-2mm]
{{{\gamma }'}^{{m}'}_{\,\,\mu {n}'}}({x}')&=[L_{\ \,k}^{{{m}'}}(x)L_{{{n}'}}^{\ \ l}(x)\gamma^{k} _{\,\,\nu l}(x)+L_{\ \,k}^{{{m}'}}(x){{\partial }_{\nu }}L_{{{n}'}}^{\ \ k}(x)]{{{\partial }'}_{\mu }}{{x}^{\nu }}
\notag
\end{align}
where $L_{\ \,k}^{{{m}'}}$ and $L_{{{n}'}}^{\ \ k}$ are mutually inverse matrices $L_{\ \,k}^{{{m}'}}L_{{{n}'}}^{\ \ k}=\delta _{{{n}'}}^{{{m}'}}$ which are distinguished from each other by the order of indexes.

Note that transformations from $S_{M}^{g}$ are considered here from an active point of view, that is, as real local linear transformations of tangent spaces $T{{M}_{x}}$  (rather than their reparameterization) and real displacements of points in the space-time manifold $M$ in fixed coordinates (rather than changing coordinates).

\subsection{Generalized principle of relativity}

In this subsection we study the consequences of $G{{L}^{g}}$-invariance of GRAF. Here we show that \textit{the Palatini equation is the strong Noether identity which follows from $G{{L}^{g}}$-invariance of GRAF.}

A distinctive feature of GRAF is the interpretation of reference frames as anholonomic in the general case affine frame fields \textit{(generalized reference frames)}, which is broader than in the EG. This extension of the concept of reference frames in GRAF leads to the extension of the general principle of relativity, which in the framework of GRAF is interpreted as the invariance of gravitational field equations with respect to the transitions between the generalized reference frames, which form the group $G{{L}^{g}}$ of local linear transformations of affine frame fields. The group $G{{L}^{g}}$ is broader than the group ${{H}^{g}}$ of diffeomorphisms of Lagrangian coordinates describing transitions between holonomic frame fields accepted in the EG as reference frames, or than the group ${{O}^{g}}$ of local Lorentz transformations describing transitions between orthonormal frame fields accepted as reference frames in GROF. Both of these groups are subgroups of the group $G{{L}^{g}}$, so the symmetry of GRAF is wider than the symmetry of EG or GROF.

\textit{The generalized principle of (general) relativity states that the gravitational field equations must have the same appearance in all generalized reference frames (affine frames), that is, be $G{{L}^{g}}$-invariant.}

Consider the $G{{L}^{g}}$-transformations $L_{\ \,k}^{m}\approx \delta _{\,k}^{m}+l_{\ \,\,k}^{m}$ with infinitesimal parameters $l_{\ \,\,k}^{m}$. In this case formulas (\ref{eq42}) take the form
\begin{align} \label{eq45}
{{\delta }_{l}}h_{\mu }^{m}&=l_{\ \,\,k}^{m}h_{\mu }^{k}, \qquad {{\delta }_{l}}{{g}_{mn}}=-l_{\ \,m}^{k}\,{{g}_{kn}}\,-l_{\ \,n}^{k}{{g}_{mk}}
\notag\\[-2mm] &\\[-2mm]
{{\delta }_{l}}\gamma _{\,\,\mu n}^{m}&=l_{\ \,k}^{m}\gamma _{\,\,\mu n}^{k}-l_{\ \,n}^{k}\gamma _{\,\,\mu k}^{m}-{{\partial }_{\mu }}l_{\ \,n}^{m}.
\notag
\end{align}
From these formulas, taking into account (\ref{eq3}) and (\ref{eq26}), it follows
\begin{equation}\label{eq47}
{{\delta }_{l}}V_{\gamma }^{\sigma }=-\Sigma _{m}^{\ \ n\nu \sigma }{{\partial }_{\nu }}l_{\ \,n}^{m}.
\end{equation}
Here ${{\delta }_{l}}$ is the variation of the shape of fields at infinitesimal $G{{L}^{g}}$-transformations, and, due to ${{\delta }_{l}}{{x}^{\mu }}=0$, ${{{\delta }'}_{l}}={{\delta }_{l}}$.

The transformations (\ref{eq8}) for $g^a\sim {l^m}_n$ take the form
$$
\begin{array}{rcl}
{{\delta }_{l}}h_{\mu }^{m} = a_{\mu n}^{m \,k}\,l_{\ \,k}^{n}, \qquad
{{\delta }_{l}}{{g}_{mn}}= a_{mnp}^{\ \quad \ k}\ l_{\ \,k}^{p}, \ \qquad
{{\delta }_{l}}{{V}^{\sigma }}= d_{m}^{\ \ n\nu \sigma }{{\partial }_{\nu }}\,l_{\ \,n}^{m}
\end{array}
$$
where in our case of transformations (\ref{eq45}), (\ref{eq47})
\begin{equation}\label{eq48}
a_{\mu n}^{m \,k}=\delta _{n}^{m}h_{\mu }^{k},\qquad
 a_{mnp}^{\ \quad \ k}=-{{g}_{pm}}\,\delta _{n}^{k}-{{g}_{pn}}\,\delta _{m}^{k},\qquad
  d_{m}^{\ \ n\nu \sigma }=-\Sigma _{m}^{\ \ n\nu \sigma }
\end{equation}
and all the relevant coefficients $b_{\,\,a}^{i\mu }=0$ and $c_{a}^{\sigma }=0$.

From the definitions (\ref{eq10}), taking into account (\ref{eq34}),  subsequent expressions for the Noether currents and their superpotentials related to $G{{L}^{g}}$-transformations follow
$$
\begin{array}{rcl}
J_{m}^{\ \ n\mu }=-B_{s}^{\sigma \mu }a_{\sigma m}^{s \, \;\, n}-\tfrac{1}{2}{{D}^{ps\mu }}a_{psm}^{\ \quad n}, \qquad
S_{m}^{\ \ n\mu \nu} =-d_{m}^{\ \ n\mu \nu }
\end{array}
$$
which with the help of (\ref{eq48}) are specified as follows
\begin{equation}\label{eq49}
J_{m}^{\ \ n\mu }=D_{m}^{\,\cdot \,\,n\,\mu }-B_{m}^{n\mu }=M_{m}^{\ \ n\mu }, \qquad
S_{m}^{\ \ n\mu \nu }=\Sigma _{m}^{\ \ n\mu \nu }.
\end{equation}

The partition (\ref{eq23}) of the Hilbert's Lagrangian $\Lambda $ into the volume ${{L}_{\gamma }}$ and the surface ${{\partial }_{\sigma }}V_{\gamma }^{\sigma }$ parts is not $G{{L}^{g}}$-invariant and when changing a general reference frame between them there is redistribution with conservation
\begin{equation}\label{eq50}
{{{\delta }'}_{l}}\Lambda ={{{\delta }'}_{l}}{{L}_{\gamma }}+{{\partial }_{\sigma }}{{{\delta }'}_{l}}V_{\gamma }^{\sigma }=0.
\end{equation}
Since for $G{{L}^{g}}$-transformations ${{\delta }_{l}}{{x}^{\mu }}=0$ and ${{\delta }_{l}}{{g}_{\mu \nu }}=0$, from (\ref{eq38}), taking into account the expressions (\ref{eq45}), it follows:
${{\delta }_{l}}{{L}_{\gamma }}=-{{\partial }_{\sigma }}(M_{m}^{\ \ n\sigma }l_{\ \,\,n}^{m})$,
which together with (\ref{eq47}) and (\ref{eq50}), gives the $G{{L}^{g}}$-invariance condition of the Hilbert's Lagrangian in (pseudo)Riemannian space
\begin{equation}\label{eq51}
{{\delta }_{l}}\Lambda ={{\partial }_{\sigma }}(\Sigma _{m}^{\ \ n\nu \sigma }{{\delta }_{l}}\gamma _{\,\,\nu n}^{m})=-{{\partial }_{\sigma }}(M_{m}^{\ \ n\sigma }l_{\ \,\,n}^{m}+\Sigma _{m}^{\ \ n\nu \sigma }{{\partial }_{\nu }}l_{\ \,n}^{m})=0.
\end{equation}

It is interesting to note that due to the fact that ${{\delta }_{l}}{{g}_{\mu \nu }}=0$, in expression (\ref{eq51}) there are no variational derivatives, so all the identities obtained from (\ref{eq51}) are strong
\begin{equation}\label{eq52}
{{\partial }_{\sigma }}M_{m}^{\ \ n\sigma }=0,\qquad
{{\partial }_{\sigma }}\Sigma _{m}^{\ \ n\nu \sigma }+M_{m}^{\ \ n\nu }=0,\qquad
\Sigma _{m}^{\ \ n\nu \sigma }=-\Sigma _{m}^{\ \ n\sigma \nu }.
\end{equation}
The second identity (\ref{eq52}) coincides with the Palatini equation (\ref{eq28}), which, as we can see, is a consequence of $G{{L}^{g}}$-invariance of the theory and is an expression of the corresponding Noether current through the superpotential.

Condition (\ref{eq51}) of $G{{L}^{g}}$-invariance of GRAF for contravariant parameters $l_{\ \ \ \cdot }^{mn}$ of group $G{{L}^{g}}$ takes the form
\begin{equation}\label{eq55}
{{\delta }_{l}}\Lambda =-{{\partial }_{\sigma }}(N_{mn}^{\ \ \ \ \sigma }l_{\ \,\,\ \cdot }^{mn}+\Sigma _{mn}^{\ \ \ \,\cdot \,\nu \sigma }{{\partial }_{\nu }}l_{\ \,\,\ \cdot }^{mn})=0
\end{equation}
where $N_{mn}^{\ \ \ \ \sigma }:=M_{mn}^{\ \ \,\,\cdot \,\sigma }+\Sigma _{m}^{\ \ s\nu \sigma }{{\partial }_{\nu }}{{g}_{sn}}$. So the Palatini equation can be represented as
\begin{equation}\label{eq56}
{{\partial }_{\sigma }}\Sigma _{mn}^{\ \ \ \,\cdot \,\nu \sigma }+N_{mn}^{\ \ \ \ \nu }=0.
\end{equation}

By the metricity condition (\ref{eq19}) we obtain
$$
\begin{array}{rcl}
N_{mn}^{\ \ \ \ \sigma }=\Sigma _{ms}^{\ \ \,\cdot \,\nu \sigma }\gamma _{\ \ \nu n}^{s}-\Sigma _{ns}^{\ \ \cdot \,\nu \sigma }\gamma _{\ \ \nu m}^{s}
\end{array}
$$
hence $N_{mn}^{\ \ \ \ \sigma }$, like $\Sigma _{mn}^{\ \ \ \,\cdot \,\nu \sigma }$, are antisymmetric in the lower indices. In this case, condition (\ref{eq55}) of the $G{{L}^{g}}$-invariance of GRAF includes only the antisymmetric components ${{\omega }^{mn}}:=l_{\ \ \ \ \,\cdot }^{[mn]}$ of parameters of the group $G{{L}^{g}}$. The transformations with parameters ${{\omega }^{mn}}$, according to (\ref{eq45}), do not change the metric in the affine frame ${{\delta }_{\omega }}{{g}_{mn}}=-{{\omega }_{mn}}\,-{{\omega }_{nm}}=0$, and therefore correspond to local (pseudo)rotations. Thus, current $N_{mn}^{\ \ \ \ \nu }$ has the physical meaning of the tensor density of the internal angular momentum (spin) of the gravitational field, and $\Sigma _{mn}^{\ \ \,\cdot \,\nu \sigma }$, according to (\ref{eq56}), has the physical meaning of its superpotential. Given also the non-torsionity condition (\ref{eq18}) we obtain
$$
\begin{array}{rcl}
N_{mn}^{\ \ \ \ \sigma }=\tfrac{e}{2}(F_{mn}^{\sigma }-h_{m}^{\sigma }{{\tilde{R}}_{n}}+h_{n}^{\sigma }{{\tilde{R}}_{m}}).
\end{array}
$$

	Transformations with symmetric components ${{\sigma }^{mn}}:=l_{\ \ \ \ \,\cdot }^{\{mn\}}$ of parameters of the group $G{{L}^{g}}$ lead to deformations of the metric in the affine frame ${{\delta }_{\sigma }}{{g}_{mn}}=-2{{\sigma }_{mn}}$ with the infinitesimal strain tensor ${{\sigma }_{mn}}$. The Noether current
\begin{equation}\label{eq57}
J_{\{mn\}}^{\ \ \ \ \sigma }=M_{\{mn\}}^{\ \ \,\,\ \,\cdot \,\sigma }+{{\partial }_{\nu }}{{g}_{s\{m}}\Sigma _{n\}}^{\ \ s\nu \sigma }
\end{equation}
which is connected with ${{\sigma }^{mn}}$, in the case of metricity (\ref{eq19}) equals zero. This is another argument in favor of the possibility of arbitrary choice of vacuum (with metric ${{g}_{mn}}$) on the basis of which gravitational phenomena are considered. This additional possibility may be useful when looking for new solutions in gravity theory.

\subsection{Gauge translational invariance}

	A gravitational field is born by an energy-momentum which associated with space-time translations. So gravity is a gauge theory of the translations group. This, regardless of the particular type of Lagrangian, determines the structure of theory. In particular, it permits to \textit{represent the gravitational field equations in superpotential form}, that is, in the form that analogous to the form of Maxwell dynamic equations.

	In this subsection we study the consequences of $T_{M}^{g}$-invariance of GRAF.

	From formulas (\ref{eq42}) follows, that $T_{M}^{g}$-transformations with infinitesimal parameters ${{\delta }_{t}}{{x}^{\mu }}={{t}^{\mu }}$ are
\begin{align} \label{eq58}
{{\delta }_{t}}h_{\mu }^{m}&=-{{t}^{\sigma }}{{\partial }_{\sigma }}h_{\mu }^{m}-h_{\sigma }^{m}{{\partial }_{\mu }}{{t}^{\sigma }}, \qquad
{{\delta }_{t}}{{g}_{mn}}=-{{t}^{\sigma }}{{\partial }_{\sigma }}{{g}_{mn}}
\notag\\[-2mm] &\\[-2mm]
{{\delta }_{t}}\gamma _{\,\,\mu n}^{m}&=-{{t}^{\sigma }}{{\partial }_{\sigma }}\gamma _{\,\,\mu n}^{m}-\gamma _{\,\,\sigma n}^{m}{{\partial }_{\mu }}{{t}^{\sigma }}.
\notag
\end{align}

The partition (\ref{eq23}) of the Hilbert's Lagrangian $\Lambda $ is $T_{M}^{g}$-invariant, because ${{{\delta }'}_{t}}{{L}_{\gamma }}=0$ and ${{{\delta }'}_{t}}V_{\gamma }^{\sigma }=0$. Therefore, the surface addend of the Hilbert's Lagrangian is not affected on the consequences of $T_{M}^{g}$-invariance of the GRAF equations.
The parameters ${{t}^{\mu }}$ describe the infinitesimal displacements along the Eulerian coordinates. They correspond to displacements $\,{{t}^{m}}=h_{\mu }^{m}\,{{t}^{\mu }}$ along the affine basis vectors that parameterize the deformed gauge group of translations  $T_{M}^{gh}$ \cite{SamR}.
Transformations (\ref{eq58}) in terms of parameters $\,{{t}^{m}}$ of the group $T_{M}^{gh}$ take the form
\begin{equation}\label{eq60}
 {{\delta }_{t}}h_{\mu }^{m}=-F_{\mu \,n}^{m}\,{{t}^{n}}-{{\partial }_{\mu }}{{t}^{m}},\qquad
 {{\delta }_{t}}{{g}_{mn}}=-{{\partial }_{s}}{{g}_{mn}}\,{{t}^{s}}.
\end{equation}
Transformations (\ref{eq8}) for ${{g}^{a}}\sim{{t}^{m}}$ give
$$
\begin{array}{rcl}
{{\delta }_{t}}h_{\mu }^{m}=a_{\mu n}^{m}\,{{t}^{n}}+b_{\mu n}^{m \nu}{{\partial }_{\nu }}{{t}^{n}},\qquad
{{\delta }_{t}}{{g}_{mn}}=a_{mns}^{\ \,}{{t}^{s}}.
\end{array}
$$
So for $T_{M}^{gh}$- transformations (\ref{eq60}) we obtain
\begin{equation}\label{eq61}
a_{\mu n}^{m}=-F_{\mu n}^{m}, \qquad
a_{mns}^{\ \,}=-{{\partial }_{s}}{{g}_{mn}}, \qquad
b_{\mu n}^{m \nu}=-\delta _{n}^{m}\delta _{\mu }^{\nu }
\end{equation}
and all other coefficients (except $h_{m}^{\mu }$) equal zero. From the definitions (\ref{eq10}), taking into account (\ref{eq19}), (\ref{eq34}), (\ref{eq61}), follow the subsequent expressions for the Noether current $t_{m}^{\mu }$ and superpotential
$S_{m}^{\mu \nu }$ related to $T_{M}^{gh}$-transformations
\begin{equation}\label{eq62}
t_{m}^{\mu }=B_{s}^{\sigma \mu }F_{\sigma m}^{s}\,+D_{s}^{\,\cdot \,p\mu }\gamma _{\,\,mp}^{s}\,-{{L}_{\gamma }}h_{m}^{\mu }, \qquad  S_{m}^{\mu \nu }=B_{m}^{\mu \nu }.
\end{equation}
Therefore, $t_{m}^{\mu }$ determines the \textit{energy-momentum of the gravitational field} in the affine frame ${{h}_{m}}$. The current $t_{m}^{\mu }$ (as well as the superpotential $B_{m}^{\mu \nu }$) is a mixed coordinate-frame tensor density with respect to the gauge translations (at which diffeomorphisms of Eulerian coordinates occur) and global linear transformations of the frame fields. In the case of gauge linear transformations of frame fields, the current $t_{m}^{\mu }$ (as well as the superpotential $B_{m}^{\mu \nu }$) is transformed by the non-tensor law \cite{Obukhov}
\begin{align} \label{eq63}
{{t}'}_{{m}'}^{\mu }&=L_{{{m}'}}^{\ \ \ m}(t_{m}^{\mu }-{{\partial }_{\sigma }}\Delta B_{m}^{\mu \sigma })-{{\partial }_{\sigma }}L_{{{m}'}}^{\ \ \ m}(B_{m}^{\mu \sigma }+\Delta B_{m}^{\mu \sigma })
\notag\\[-2mm] &\\[-2mm]
{{B}'}_{{m}'}^{\mu \sigma }&=L_{{{m}'}}^{\ \ \ m}(B_{m}^{\mu \sigma }+\Delta B_{m}^{\mu \sigma })
\notag
\end{align}
where $\Delta B_{m}^{\mu \sigma }:=e\,(\Delta \gamma _{\ \ \,m\ .}^{[\mu \ \sigma ]}-h_{m}^{\mu }\,\Delta \gamma _{\,\,n\,\ \cdot }^{[n\,\sigma ]}+h_{m}^{\sigma }\,\Delta \gamma _{\,\,n\,\ \cdot }^{[n\,\mu ]})$, $\Delta \gamma _{\,\,\nu n}^{m}:={{\partial }_{\nu }}L_{{{s}'}}^{\ \ m}L_{\ \,n}^{{{s}'}}$. This is due to the extra energy-momentum that must be transmitted for the transition between mutually accelerated reference frames \cite{Maluf}.

Note that to denote quantities having different tensor dimensions, one letter can be used without danger of confusing these quantities, such as the translation parameters ${{t}^{m}}$ and the tensor density $t_{m}^{\sigma }$ of the energy-momentum of the gravitational field.

Taking into account (\ref{eq38}) and (\ref{eq60}), the $T_{M}^{gh}$-invariance condition of GRAF is written as
\begin{equation}\label{eq65}
{{{\delta }'}_{t}}\Lambda =-eG_{n}^{\nu }(\,\gamma _{\,\,\nu m}^{n}\,{{t}^{m}}+{{\partial }_{\nu }}{{t}^{n}})-{{\partial }_{\nu }}(t_{m}^{\nu }\,{{t}^{m}}+B_{m}^{\mu \nu }{{\partial }_{\mu }}{{t}^{m}})=0
\end{equation}
so we obtain the following strong Noether identities
\begin{equation}\label{eq66}
-eG_{n}^{\nu }\,\gamma _{\nu m}^{n}={{\partial }_{\nu }}t_{m}^{\nu },\qquad
-eG_{m}^{\mu }={{\partial }_{\nu }}B_{m}^{\mu \nu }+t_{m}^{\mu },\qquad
B_{m}^{\mu \nu }=-B_{m}^{\nu \mu }.
\end{equation}

The second identity (\ref{eq66}) can be represented in the form (\ref{eq15}), which in our case gives
$$
\begin{array}{rcl}
t_{m}^{\nu }=-{{\partial }_{h_{\nu }^{m}}}{{L}_{\gamma }}.
\end{array}
$$
	
In the presence of matter fields with the tensor density $\tau _{m}^{\mu }$ of the energy-momentum, when Einstein equation $eG_{m}^{\mu }=\tau _{m}^{\mu }$ are fulfilled, the second identity (\ref{eq66}) takes the form
\begin{equation}\label{eq69}
{{\partial }_{\nu }}B_{m}^{\mu \nu}=-T_{m}^{\mu}
\end{equation}
where $T_{m}^{\mu }=t_{m}^{\mu }\,+\tau _{m}^{\mu }$ is the total tensor density of the energy-momentum of the gravitational and matter fields. Conversely, from the equation (\ref{eq69}), given second identity (\ref{eq66}), the Einstein equation follows. Therefore, \textit{equation} (\ref{eq69}) \textit{is the Einstein equation in the affine frame}, which is written in the superpotential form, that is, in the form similar to that of Maxwell's dynamic equations (and equations in other gauge theories). The tensor density $B_{m}^{\mu \nu }$, called the \textit{induction of the gravitational field} in the GRAF, acts as a superpotential of the total energy-momentum $T_{m}^{\mu }$ of the gravitational and matter fields.

Representation of the Einstein equation in the form (\ref{eq69}), similar to the form of field equations of any other gauge theory, is a consequence of $T_{M}^{gh}$-symmetry of the theory of gravity. This is convincingly illustrated the fact that gravity, as an interaction generated by energy-momentum, is the gauge theory of the translation group, and the superpotential of the total energy-momentum $T_{m}^{\mu}$ (induction of the gravitational field $B_{m}^{\mu \nu}$) acts as the power characteristic of the gravitational field (but not a curvature, as in EG, or a torsion, as in TEGR). Such an interpretation opens new avenues for unification of gravity with other gauge theories by uniting appropriate generalized deformed gauge groups \cite{Sam5}, which may be nontrivial \cite{Sam6}.

From the equation (\ref{eq69}), considering (\ref{eq66}), the conservation law follows
\begin{equation}\label{eq70}
{{\partial }_{\mu }}T_{m}^{\mu }=0.
\end{equation}

On the gravitational extremal $eG_{m}^{\mu }=\tau _{m}^{\mu }$, taking into account (\ref{eq70}) ${{\partial }_{\sigma }}t_{m}^{\sigma }=-{{\partial }_{\sigma }}\tau _{m}^{\sigma }$, first identity (\ref{eq66}) goes into the equation of motion of (macro)matter
\begin{equation}\label{eq71}
{{\partial }_{\mu }}\tau _{m}^{\mu }\,=\tau _{s}^{\mu }\gamma _{\mu m}^{s}.
\end{equation}
For the dust matter $\tau _{m}^{\mu }\,=e \mu \,{{u}^{\mu }}{{u}_{m}}$, where $\mu $ is the density of its mass, $u_{{}}^{\mu }\,={{\dot{x}}^{\mu }}$. Provided that the mass is conserved ${{\partial }_{\mu }}(e \mu {{u}}^{\mu })=0$ (that is, in the absence of reactive forces), equation (\ref{eq71}) is the equation of geodesic $\dot{u}_{{}}^{m}+\gamma _{\,sn}^{m}u_{{}}^{s}u_{{}}^{n}=0$ on which the particles move.

\subsection{Complete symmetry of the general relativity}

	Under energy-momentum we understand a Noether current which is conserved as a result of the space-time translational invariance. But translations can simultaneously cause $G{{L}^{g}}$-transformations of the reference frame (as in the case of EG), which lead to the energy-momentum renormalization. This situation corresponds to the deformation of the group $S_{M}^{gh}\to S_{M}^{gH}$ \cite{Sam4}. The multiplicity of deformed specimens of the group $S_{M}^{gH}$ leads to the multiplicity of variants of Noether currents which are conserved due to translational invariance of GRAF, hence leading to a \textit{plurality of ways of determining the energy-momentum of the gravitational field and the corresponding superpotential.} Among all deformed groups $S_{M}^{gH}$ there is one highlighted, so-called group of parallel transports $DP$, transformations of which are consistent with the geometric structure of a (pseudo)Riemannian space \cite{Sam4}. This group corresponds to the energy-momentum tensor density, which is characterized by its dependence of the derivatives of connection coefficients, so the energy-momentum in this case is localizable.

	In this subsection, we explore these issues.

	Consider the infinitesimal variant of transformations of a complete symmetry group $S_{M}^{gh}=G{{L}^{g}}\times\!\!\!|\, T_{M}^{gh}$, considering both transformations (\ref{eq45}) of the subgroup $G{{L}^{g}}$ and transformations (\ref{eq60}) of the subgroup $T_{M}^{gh}$ simultaneously ${{\delta }_{g}}={{\delta }_{l}}+{{\delta }_{t}}$
\begin{equation}\label{eq72}
{{\delta }_{g}}h_{\mu }^{m}=l_{\ \,\,k}^{m}h_{\mu }^{k}-F_{\mu \,n}^{m}\,{{t}^{n}}-{{\partial }_{\mu }}{{t}^{m}}, \quad {{\delta }_{g}}{{g}_{mn}}=-l_{\ \,m}^{k}\,{{g}_{kn}}\,-l_{\ \,n}^{k}{{g}_{mk}}-{{\partial }_{s}}{{g}_{mn}}\,{{t}^{s}}.
\end{equation}

The $S_{M}^{gh}$-invariance condition of GRAF is obtained by combining conditions (\ref{eq51}) and (\ref{eq65})
\begin{align} \label{eq75}
{{{\delta }'}_{g}}\Lambda =&-eG_{n}^{\nu }(\,\gamma _{\,\,\nu m}^{n}\,{{t}^{m}}+{{\partial }_{\nu }}{{t}^{n}})
\notag\\[-2mm] &\\[-2mm]
&-{{\partial }_{\sigma }}(t_{m}^{\sigma }\,{{t}^{m}}+B_{m}^{\nu \sigma }{{\partial }_{\nu }}{{t}^{m}}+M_{m}^{\ \ n\sigma }l_{\ \,\,n}^{m}+\Sigma _{m}^{\ \ n\nu \sigma }{{\partial }_{\nu }}l_{\ \,n}^{m})=0.
\notag
\end{align}

We deform the group $S_{M}^{gh}\to S_{M}^{gH}$, assuming that an additional linear transformation $dl_{\ \,\,n}^{m}=-\lambda _{s\,n}^{m}\,{{t}^{s}}$ occurs at the translations, that is,
introducing instead of $l_{\ \,\,n}^{m}$ the new parameters of infinitesimal linear transformations
\begin{equation}\label{eq76}
r_{\ \,\,n}^{m}=l_{\ \,\,n}^{m}+\lambda _{s\,n}^{m}\,{{t}^{s}}
\end{equation}
with arbitrary functions $\lambda _{s\,n}^{m}$. In this parameterization (that is, for the group $S_{M}^{gH}$), transformations (\ref{eq72}) take the form
\begin{align} \label{eq77}
{{\delta }_{g}}h_{\mu }^{m}&=r_{\ \,\,k}^{m}h_{\mu }^{k}-(F_{\mu \,n}^{m}+\lambda _{n\,\mu }^{m})\,{{t}^{n}}\,-{{\partial }_{\mu }}{{t}^{m}}
\notag\\[-2mm] &\\[-2mm]
{{\delta }_{g}}{{g}_{mn}}&=-r_{\ \,m}^{k}\,{{g}_{kn}}\,-r_{\ \,n}^{k}{{g}_{mk}}-({{\partial }_{s}}{{g}_{mn}}-\lambda _{msn}^{\cdot }-\lambda _{nsm}^{\cdot })\,{{t}^{s}}.
\notag
\end{align}
The deformation of the group $S_{M}^{gh}$ changes the way of forming the semidirect product of the groups $G{{L}^{g}}$ and $T_{M}^{gh}$, which is determined by the transformations (\ref{eq77})  \cite{Sam4}.

The $S_{M}^{gH}$-invariance condition of the GRAF can be written in the form
\begin{align}
{{{\delta }'}_{g}}\Lambda =&-eG_{n}^{\nu }(\,\gamma _{\,\,\nu m}^{n}\,{{t}^{m}}+{{\partial }_{\nu }}{{t}^{n}})\notag\\
&-{{\partial }_{\sigma }}\left( \left( t_{m}^{\sigma }+{{\partial }_{\nu }}\left( \,e\lambda _{\ \! \ m\,\cdot }^{[\sigma \ \,\nu ]} \right) \right)\,\,{{t}^{m}}+\left( B_{m}^{\nu \sigma }-e\lambda _{\ \! \ m\,\cdot }^{[\nu \ \,\sigma ]} \right)\,\,{{\partial }_{\nu }}{{t}^{m}}\right)\label{eq79}\\
&+{{\partial }_{\sigma }}\left(M_{m}^{\ \ n\sigma }r_{\ \,\,n}^{m}+\Sigma _{m}^{\ \ n\nu \sigma }{{\partial }_{\nu }}r_{\ \,n}^{m} \right)=0\notag
\end{align}
which follows from condition (\ref{eq75}), taking into account (\ref{eq76}) and the Palatini equation (\ref{eq28}). Therefore, during the deformation of the group $S_{M}^{gh}$, the energy- momentum of the gravitational field and correspondent superpotential are renormalized
\begin{align} \label{eq80}
t_{m}^{\nu }\to J_{m}^{\nu }&=t_{m}^{\nu }+{{\partial }_{\sigma }}\left( \,e\lambda _{\ \! \ m\,\cdot }^{[\nu \ \,\sigma ]}\right),\qquad B_{m}^{\nu \sigma }\to S_{m}^{\nu \sigma }=B_{m}^{\nu \sigma }-e\lambda _{\ \! \ m\,\cdot }^{[\nu \ \,\sigma ]}
\notag\\[-2mm] &\\[-2mm]
S_{m}^{\nu \sigma }&=e\gamma _{\ \! \ m\ \! \cdot }^{[\nu \ \sigma ]}-e\lambda _{\ \! \ m\ \! \cdot }^{[\nu \ \sigma ]}-h_{m}^{\nu }\,{{V}_{\gamma}^{\sigma }}+h_{m}^{\sigma }\,{{V}_{\gamma}^{\nu }}.
\notag
\end{align}

Taking into account second identity (\ref{eq66}), we see that the deformation of the group $S_{M}^{gh}$ leads to the redistribution between the divergence of superpotential and the tensor density of the energy-momentum of the gravitational field in the Einstein tensor
$$
\begin{array}{rcl}
-eG_{m}^{\mu }={{\partial }_{\nu }}B_{m}^{\mu \nu }+t_{m}^{\mu }\,={{\partial }_{\nu }}S_{m}^{\mu \nu }+J_{m}^{\mu }
\end{array}
$$
illustrating, in the case of the group $S_{M}^{gH}$, the general position that take place for generalized deformed gauge groups and which was indicated by formulas (\ref{eq16}) and the text after them.

Special choice of $\lambda _{s\,n}^{m}$ allows to discover a lot of "new internal gauge symmetries" in the theory of gravity \cite{Mont}. Now let
\begin{equation}\label{eq82}
\lambda _{s\,n}^{m}=\gamma _{s\,n}^{m}.
\end{equation}
In this case, given the non-torsionity (\ref{eq18}) and metricity (\ref{eq19}) conditions, transformations (\ref{eq77}) are greatly simplified
\begin{equation}\label{eq83}
{{\delta }_{g}}h_{\mu }^{m}=r_{\ \,\,k}^{m}h_{\mu }^{k}-{{D}_{\mu }}{{t}^{m}},\qquad
{{\delta }_{g}}{{g}_{mn}}=-r_{\ \,m}^{k}\,{{g}_{kn}}\,-r_{\ \,n}^{k}{{g}_{mk}}
\end{equation}
where ${{D}_{\mu }}{{t}^{m}}={{\partial }_{\mu }}{{t}^{m}}+\gamma _{\mu \,n}^{m}\,{{t}^{n}}$ is the covariant derivative of the vector ${{t}^{m}}$. Note that in pure translations from the group $S_{M}^{gH}$ (at $r_{\ \,\,k}^{m}=0$) the metric in the affine frame does not change
$$
\begin{array}{rcl}
{{\delta }_{\,t}}{{g}_{mn}}=0.
\end{array}
$$
In addition, from (\ref{eq83}) in this case we obtain
$$
\begin{array}{rcl}
{{{h}'}_{m}}({x}')={{h}_{m}}(x)+{{t}^{s}}\gamma _{s\,m}^{n}{{h}_{n}}(x),
\end{array}
$$
which corresponds to the definition of the affine connection by the method of Cartan's moving frame. The group $S_{M}^{gH}$, provided (\ref{eq82}), is the group $DP$ of parallel transports in the Riemannian space \cite{Sam4}. Transformations from the $DP$ set on the $M$ the structure of the (pseudo)Riemannian space with the metric ${{g}_{mn}}$ in the affine frame ${{h}_{m}}$.
Note that translations do not form a subgroup in $DP$ since $[{{D}_{\mu }}{{D}_{\mu }}]_{\ n}^{m}=R_{\ n\mu \nu }^{m}\in AlG{{L}^{g}}$.

From (\ref{eq80}) it follows that for $DP$
\begin{equation}\label{eq85}
J_{m}^{\nu }=t_{m}^{\nu }+e{{D}_{\sigma }}\gamma _{\ \ m\,\cdot }^{[\nu \ \,\sigma ]},\qquad
S_{m}^{\nu \sigma }=-h_{m}^{\nu }\,{{V}_{\gamma}^{\sigma }}+h_{m}^{\sigma }\,{{V}_{\gamma}^{\nu }}=-\delta _{m\mu }^{\nu \,\sigma }{{V}_{\gamma}^{\mu }}
\end{equation}
where ${{D}_{\sigma }}\gamma _{\ \ m\,\cdot }^{[\nu \ \,\sigma ]}=\frac{1}{e}{{\partial }_{\sigma }}\left( e\gamma _{\ \ m\,\cdot }^{[\nu \ \,\sigma ]} \right)$ is the covariant divergence of the antisymmetric $T_{M}^{g}$-tensor $\gamma _{\ \ m\,\cdot }^{[\nu \ \,\sigma ]}$. It is interesting to note that in case of group $DP$,  superpotential  $S_{m}^{\nu \sigma }$ depends solely of vector density ${{V}_{\gamma}^{\mu }}$ , that determines the surface part of Hilbert's Lagrangian $\Lambda $. In addition, in expression (\ref{eq85}) for $J_{m}^{\nu }$ there are derivatives of $\gamma _{\ \ m\,\cdot }^{[\nu \ \,\sigma ]}$  , so the energy-momentum of the gravitational field in this case is localizable.

\section{Special Gauge Conditions}

The broad symmetry of GRAF is provided by a sufficient number of field variables of the theory. Limiting such symmetry by imposing gauge conditions reduces the number of field variables. \textit{Different variants of the gauge conditions lead to different locally equivalent formulations of the general relativity,} the most characteristic of which we consider in this section.

Keeping the status of gravity as a gauge theory of a translation group, we will not consider the gauge conditions that violate the $T_{M}^{g}$-invariance of the theory. We restrict ourselves to the gauge conditions that narrow the $G{{L}^{g}}$-invariance of GRAF, which corresponds to the choice of special classes of generalized reference frames (affine frames).

\subsection{Einstein gravity}

	From the outset, general relativity was formulated by Einstein in holonomic reference frames, which are described by frame fields with zero anholonomic objects, and can therefore be implemented as coordinate frame fields of certain coordinate systems. In this case, the gauge condition that limits the $G{{L}^{g}}$-invariance of the theory is the holonomicity condition
	\begin{equation}\label{eq87}
F_{\mu \nu }^{m}={{\partial }_{\nu }}h_{\mu }^{m}-{{\partial }_{\mu }}h_{\nu }^{m}=0
\end{equation}
which ensures that there are functions ${{y}^{m}}(x)$ such that $h_{\mu }^{m}={{\partial }_{\mu }}{{y}^{m}}$. The quantities ${{y}^{m}}$ which  give holonomic reference frames play the role of Lagrangian coordinates, in contrast to the coordinates ${{x}^{\mu }}$ in space-time, which numerate its points and play the role of Eulerian coordinates. Therefore, the functions ${{y}^{m}}(x)$ are scalars with respect to the transformations of the coordinates ${{{x}'}^{\mu }}={{{x}'}^{\mu }}(x)$, that is, invariant under the gauge translations of the space-time ${{{x}'}^{\mu }}={{x}^{\mu }}+\delta {{x}^{\mu }}(x)$: ${{{y}'}^{m}}({x}')={{y}^{m}}(x)$, which, in fact, allows them to be interpreted as Lagrangian coordinates of the gravitational system. The vectors ${{h}_{m}}$ of holonomic frames are the coordinate vectors ${{\partial }_{m}}$ of the coordinates ${{y}^{m}}$.
	
Unlike Eulerian coordinates ${{x}^{\mu }}$, numerating space-time points and which changes should not alter any physical quantities, changes of Lagrangian coordinates ${{y}^{m}}$ alter the physical reference frame and certain physical quantities may vary. Thus, during the transition between mutually accelerated reference frames, the energy-momentum must change. If the transformations of the world coordinates ${{x}^{\mu }}$ correspond to the gauge translations in space-time in a fixed general reference frame and belong to the group $T_{M}^{g}$, then with the changes of the Lagrangian coordinates ${{y}^{m}}$ the  translations do not occur, but transformations of holonomic reference frames take place. Such transformations belong to another group, namely the group $G{{L}^{g}}$, forming its subgroup ${{H}^{g}}$ which is isomorphic to the  group of diffeomorphisms of  space-time $T_{M}^{g}$. Despite the isomorphism of the groups ${{H}^{g}}$ and $T_{M}^{g}$, they have different physical meaning and their identification leads to known difficulties in determining the energy-momentum of the gravitational field.

The symmetry group of EG is $\bar{S}_{M}^{g}={{H}^{g}}\times\!\!\!|\, T_{M}^{g}$.
	
Due to the condition (\ref{eq87}), in holonomic reference frames $\omega _{kn}^{m}=0$, and connection coefficients $\gamma _{k\,n}^{m}$ are reduced to $\Gamma _{k\,n}^{m}=\sigma _{k\,n}^{m}$, which are Christoffel symbols in Lagrangian coordinates ${{y}^{m}}$, and thus behave as scalars when transformation of Eulerian coordinates ${{x}^{\mu }}$ occurs. As for an arbitrary affine frame, this fact provides a general-covariance (gauge translational invariance) of the partition (\ref{eq23}) of the Hilbert's Lagrangian into bulk and surface parts $\Lambda ={{L}_{\Gamma }}+{{\partial }_{\sigma }}V_{\Gamma }^{\sigma }$ \cite{Landau}, where
\begin{equation}\label{eq88}
{{L}_{\Gamma }}=\Sigma _{m}^{\ \ n\mu \nu }\Gamma _{\mu \,s}^{m}\,\Gamma _{\nu \,n}^{s}=e\Gamma _{\,\,m\,s}^{[m}\,\Gamma _{\,n\,\,\,\cdot }^{|s|\,n]},\qquad
V_{\Gamma }^{\sigma }=\Sigma _{m}^{\ \ n\nu \sigma }\,\Gamma _{\nu \,n}^{m}=e\Gamma _{\,\,n\,\,\cdot }^{[n\,\sigma ]}
\end{equation}
(vertical bars highlight indices that are skipped during antisymmetrization).

The action associated with the volume Lagrangian ${{L}_{\Gamma }}$ can be represented as an integral with respect to Lagrangian coordinates ${{y}^{m}}$
$$
\begin{array}{rcl}
{{S}_{\Gamma }}=\int{{{L}_{\Gamma }}}dx=\tfrac{1}{2}\int{\delta _{mn}^{kp}\,\Gamma _{k\,s}^{m}\,\Gamma _{p\,\,\cdot }^{s\,n}\,v\,dy}
\end{array}
$$
where $dx$ and $dy=\partial ydx$ are the coordinate volumes of the Eulerian and Lagrangian coordinates, $v:=\sqrt{|{{g}_{mn}}|}$, $\partial y:=\left| {{\partial }_{\mu }}{{y}^{m}} \right|$, $e=\partial y\,v$. So, action ${{S}_{\Gamma }}$ is obviously invariant under the transformations of the world (Eulerian) coordinates ${{x}^{\mu }}$.
	
The tensor density of the energy-momentum of the gravitational field (\ref{eq62}), by condition (\ref{eq87}), is given by
$$
\begin{array}{rcl}
t_{m}^{\mu }={{\partial }_{p}}{{x}^{\mu }}(D_{\,\,\,s}^{n\,\cdot \,p}\Gamma _{nm}^{s}\,-{{L}_{\Gamma }}\delta _{m}^{p}).
\end{array}
$$
In our case, second formula (\ref{eq36}) is reduced to
$$
\begin{array}{rcl}
D_{{}}^{ns\,p}=\partial y\,v\,(-\Gamma _{\quad \cdot \ \,\cdot }^{pns}+\,\Gamma _{k\ \,\,\,\cdot }^{k\,\{n}{{g}^{s\}p}}-{{g}^{ns}}\Gamma _{\,\,\,k\,\ \cdot }^{[k\,p]})
\end{array}
$$
therefore	
\begin{equation}\label{eq90}
t_{m}^{\mu }=\partial y\,{{\partial }_{p}}{{x}^{\mu }}\,\bar{t}_{m}^{p}
\end{equation}
where
\begin{equation}\label{eq91}
\bar{t}_{m}^{p}=v\,(-\Gamma _{sn}^{p}\Gamma _{m\,\cdot \,}^{s\,n}+\,\Gamma _{k\,n}^{k\,}\Gamma _{\ \,m\,\cdot \,}^{\{pn\}}-\Gamma _{\,\,\,n\,\ \cdot }^{[n\,p]}\Gamma _{k\,m}^{k\,}\,-\Gamma _{\,\,k\,s}^{[k}\,\Gamma _{\,\,n\,\,\cdot }^{|s|\,n]}\,\delta _{m}^{p}).
\end{equation}
	
The induction of the gravitational field (\ref{eq36}) in holonomic frames is specified as follows
\begin{equation}\label{eq92}
B_{m}^{\mu \sigma }=\partial y\,{{\partial }_{p}}{{x}^{\mu }}{{\partial }_{s}}{{x}^{\sigma }}\bar{B}_{m}^{ps}
\end{equation}
where
\begin{equation}\label{eq93}
\bar{B}_{m}^{ps}=v\,(\Gamma _{\ \ \,m\,\,\cdot }^{[p\,s]}-\delta _{m}^{p}\,\Gamma _{\,\,\,n\,\ \cdot }^{[n\,\,s]}+\delta _{m}^{s}\,\Gamma _{\,\,\,n\,\ \cdot }^{[n\,p]}).
\end{equation}
	
In relation to the transformations of Eulerian coordinates ${{x}^{\mu }}$, the energy-momentum of the gravitational field, according to (\ref{eq90}), is described by four vector densities $t_{m}^{\sigma }$, and the induction of the gravitational field given by formula (\ref{eq92}) is described by four antisymmetric tensor densities $B_{m}^{\sigma \mu }$, which correspond to four Lagrangian coordinates ${{y}^{m}}$. In the case of nonlinear transformations of Lagrangian coordinates ${{y}^{{{m}'}}}={{y}^{{{m}'}}}(y)$, both $t_{m}^{\sigma }$ and $B_{m}^{\sigma \mu }$ are transformed by non-tensor laws specifying formulas (\ref{eq63}) for our case with $L_{\ m}^{{{m}'}}={{\partial }_{m}}{{y}^{{{m}'}}}$.

In the case of identification of Lagrangian coordinates with Eulerian ones ${{y}^{m}}={{x}^{m}}$, $t_{m}^{\sigma }$ goes to the Einstein's complex $\bar{t}_{m}^{p}$ of the energy-momentum of the gravitational field, and $B_{m}^{\mu \sigma }$ to the Freud's superpotential $\bar{B}_{m}^{ps}$ \cite{Sam3} (this fact was first noted in \cite{Obukhov}). In this case, the tensor properties of $\bar{t}_{m}^{p}$ and $\bar{B}_{m}^{ps}$ at the transformations of the coordinates ${{x}^{m}}$ are violated precisely because of the  identification ${{y}^{m}}={{x}^{m}}$, which rigidly links translations in space-time with the changes of reference frames.

Einstein's equation in superpotential form ${{\partial }_{\nu }}B_{m}^{\mu \nu }=-t_{m}^{\mu }-\tau _{m}^{\mu }$ (\ref{eq69}) in holonomic frames can be represented solely in Lagrangian coordinates by means of Freud's superpotential $\bar{B}_{m}^{ps}$ and the Einstein's complex $\bar{t}_{m}^{p}$ as follows
\begin{equation}\label{eq94}
{{\partial }_{s}}\bar{B}_{m}^{ps}=-\bar{t}_{m}^{p}\,-\bar{\tau }_{m}^{p}
\end{equation}
where $\bar{\tau }_{m}^{p}=\partial {{y}^{-1}}\,{{\partial }_{\mu }}{{y}^{p}}\tau _{m}^{\mu }$ \cite{Sam3}.
	
Taking into account the fact that $\Gamma _{s\,n}^{s}=\tfrac{1}{2}{{g}^{pk}}{{\partial }_{n}}{{g}_{pk}}=\tfrac{1}{v}{{\partial }_{n}}v$, as well as condition (\ref{eq87}), in holonomic reference frames the tensor density of the internal angular momentum (spin) of the gravitational field and its superpotential can be represented as $ N_{mn}^{\quad \nu }=-\partial y{{\partial }_{[m}}{{x}^{\nu }}{{\partial }_{n]}}v $, $\Sigma _{mn}^{\ \ \ \cdot \,\nu \sigma }= \tfrac{1}{2}\partial y\,v\,\delta _{mn}^{\nu \,\sigma }$.

Since holonomic reference frames are completely determined by the functions ${{y}^{m}}={{y}^{m}}(x)$ which bind the Lagrangian and Eulerian coordinates, they can be selected as independent field variables. The Lagrangian ${{L}_{\Gamma }}$ (\ref{eq88}), like the complete Lagrangian of the gravitational system $L={{L}_{\Gamma }}+{{L}_{\psi }}$ (where ${{L}_{\psi }}$ is the Lagrangian of matter fields), does not depend on the functions ${{y}^{m}}(x)$ itself, but only on their derivatives. Therefore, the fields ${{y}^{m}}(x)$ are cyclic coordinates of the system and the equations of motion for them are reduced to the condition of preserving the corresponding generalized momentum $P_{m}^{\sigma }={{\partial }_{{{\partial }_{\sigma }}{{y}^{m}}}}L$: ${{\partial }_{\sigma }}P_{m}^{\sigma }=0$, which (up to an overall sign) coincides with the tensor density of energy-momentum of the gravitational system: $P_{m}^{\sigma }=-T_{m}^{\sigma }$.

At the end of this subsection, we consider the transformation of gauge translations along Lagrangian coordinates, which illustrates the significant physical difference between Lagrangian and Eulerian coordinates.

Gauge translations along the Lagrangian coordinates ${{\delta }_{d}}{{y}^{m}}={{d}^{m}}$ form ordinary gauge group ${{D}^{g}}=\sum\limits_{x\in M}{\ T}\subset G{{L}^{g}}$ and correspond to the transformations of the frame fields with infinitesimal parameters $l_{\,\ n}^{m}={{\partial }_{n}}{{d}^{m}}$, therefore they are given by the formulas
$$
\begin{array}{rcl}
{{\delta }_{d}}{{x}^{\mu }}=0,\qquad
{{\delta }_{d}}h_{\mu }^{m}={{\partial }_{\mu }}{{d}^{m}},\qquad
{{\delta }_{d}}{{g}_{mn}}=-2{{g}_{k\{m}}{{\partial }_{n\}}}\,{{d}^{k}}
\end{array}
$$
which follow from formulas (\ref{eq45}) with $l_{\,\ n}^{m}={{\partial }_{n}}{{d}^{m}}$ (compare with (\ref{eq60})). Invariance with respect to such transformations leads to identity
$$
\begin{array}{rcl}
{{\delta }_{d}}\Lambda =-{{\partial }_{\sigma }}(C_{m}^{\ \ \mu \sigma }{{\partial }_{\mu }}{{d}^{m}}+\Sigma _{m}^{\ \ \mu \nu \sigma }\partial _{\mu \nu }^{2}{{d}^{m}})=0
\end{array}
$$
which follows from identity (\ref{eq51}) for $l_{\,\ n}^{m}={{\partial }_{n}}{{d}^{m}}$, where $C_{m}^{\ \ \mu \sigma }=M_{m}^{\ \ \mu \sigma }+\Sigma _{m}^{\ \ n\nu \sigma }{{\partial }_{\nu }}h_{n}^{\mu }$. It follows that the current associated with ${{D}^{g}}$-transitions along the Lagrangian coordinates, like the corresponding superpotential $S_{m}^{\ \ \mu \nu }=C_{m}^{\ \ \mu \nu }+{{\partial }_{\sigma }}\Sigma _{m}^{\ \ \mu \nu \sigma }$, is equal to zero.

\subsection{General relativity in an orthonormal frame}

	The disadvantage of holonomic reference frames is their inevitable deformation in the curve space during the transition between the reference frames with the infinitesimal strain tensor ${{\sigma }^{mn}}={{g}^{s\{m}}{{\partial }_{s}}\delta {{y}^{n\}}}$. Such a drawback is obviously absent in general reference frames for which ${{g}_{mn}}=const$, in particular in (pseudo)ortho-normal reference frames, when ${{g}_{mn}}={{\eta }_{mn}}$.
In this case $\sigma _{kn}^{m}=0$, $e=h:=\left| h_{\mu }^{m} \right|$ and the connection coefficients $\gamma _{kn}^{m}$ are reduced to the quantities  $\Omega _{kn}^{m}=\omega _{kn}^{m}$, which is Ricci rotation coefficients. When we consider only orthonormal reference frames, the group $G{{L}^{g}}$ narrows to the group ${{O}^{g}}$ of gauge Lorentz transformations.

	The symmetry group of GROF is $\hat{S}_{M}^{g}=O{{}^{g}}\times\!\!\!|\, T_{M}^{g}$.

	The Lagrangian of general relativity in an orthogonal frame and the corresponding vector defining the surface term are written as \cite{Moller},  \cite{Duan}
$$
\begin{array}{rcl}
{{L}_{\Omega }}=\Sigma _{m}^{\ \ n\mu \nu }\Omega _{\mu \,s}^{m}\,\Omega _{\nu \,n}^{s}=e\,\Omega _{\,\,m\,s}^{[m}\,\Omega _{\,n\,\,\,\cdot }^{|s|\,n]},\qquad V_{\Omega }^{\sigma }=\Sigma _{m}^{\ \ n\nu \sigma }\,\Omega _{\nu \,n}^{m}=e{{R}^{\sigma }}
\end{array}
$$
where ${{R}_{n}}={{\nabla }_{\sigma }}h_{n}^{\sigma }=F_{s\,n}^{s}$ is a vector that has an important geometric meaning and determines the rate of change of the local coordinate volume along the directions of basis vectors of the selected orthonormal frame. When writing these formulas, equality $\Omega _{\,s\,n}^{s}={{R}_{n}}$ is taken into account.

The induction of the gravitational field in orthonormal reference frames is given by
\begin{equation}\label{eq95}
B_{m}^{\mu \sigma }=e\,(\Omega _{\,m\, \cdot}^{\,\mu \:\,\sigma }-h_{m}^{\mu }\,{{R}^{\sigma }}+h_{m}^{\sigma }\,{{R}^{\mu }}).
\end{equation}

A distinctive feature of orthonormal reference frames is the fact that the connection coefficients are uniquely determined by the induction of the gravitational field, that is, formula (\ref{eq95}) can be reversed. Indeed, from (\ref{eq95}) follows $B_{n}^{n\mu }=-(d-2)e\,{{R}^{\mu }}$, hence
$$
\begin{array}{rcl}
e \Omega _{\,m\, \cdot}^{\,\mu \,\,\sigma }=B_{m}^{\mu \sigma }-(h_{m}^{\mu }\,B_{n}^{n\sigma }-h_{m}^{\sigma }\,B_{n}^{n\mu })/(d-2).
\end{array}
$$

The energy-momentum tensor density of gravitational field (\ref{eq62}) in our case due to $D_{s}^{n\,\sigma }=0$ is determined by the expression \cite{Moller},  \cite{Duan}
\begin{equation}\label{eq96}
t_{m}^{\mu }=B_{s}^{n\mu }F_{n\,m}^{s}\,\,-{{L}_{\Omega }}h_{m}^{\mu }.
\end{equation}

Einstein equation in GROF specifies equation (\ref{eq69})
$$
\begin{array}{rcl}
{{\partial }_{\sigma }}B_{m}^{\mu \sigma }=-t_{m}^{\mu }-\tau _{m}^{\mu }
\end{array}
$$
where $B_{m}^{\mu \sigma }$ and $t_{m}^{\mu }$ are given by formulas (\ref{eq95}) and (\ref{eq96}) respectively. Note that in this form Einstein equation is represented in TEGR \cite{Maluf}.
	
The transformations of the deformed group $\hat{S}_{M}^{gH}=O{{}^{g}}\times\!\!\!|\, T_{M}^{g}$ under condition (\ref{eq82}) (which in our case is $\lambda _{s\,n}^{m}=\Omega _{s\,n}^{m}$) have the form
$$
\begin{array}{rcl}
{{\delta }_{g}}h_{\mu }^{m}=\omega _{\ \,\,k}^{m}h_{\mu }^{k}-{{D}_{\mu }}{{t}^{m}},\qquad
{{\delta }_{g}}{{g}_{mn}}=0
\end{array}
$$
and coincide (in the absence of torsion) with generalized Poincare gauge transformations \cite{Hehl}, \cite{Kir}. In this case, the energy-momentum of the gravitational field and the corresponding superpotential (\ref{eq85}) are given by
$$
\begin{array}{rcl}
J_{m}^{\nu }=t_{m}^{\nu }+e{{D}_{\sigma }}\Omega _{\,m\,\cdot }^{\nu \ \,\sigma },\qquad
S_{m}^{\nu \sigma }=-e\delta _{m\mu }^{\nu \,\sigma }{{R}^{\mu }}.
\end{array}
$$
	
The symmetry with respect to ${{O}^{g}}$-transformations determine the gravitational field spin, which current density and the corresponding superpotential are given by the expressions
$ N_{mn}^{\ \ \ \sigma }=\tfrac{e}{2}(F_{mn}^{\sigma }-h_{m}^{\sigma }{{R}_{n}}+h_{n}^{\sigma }{{R}_{m}})$, $ \Sigma _{mn}^{\ \ \,\cdot \,\sigma \nu }=\tfrac{e}{2}\,\delta _{mn}^{\sigma \nu }$.

\subsection{Dilaton gravity, unimodular gravity, and so on}
	
The Weyl transformations
\begin{equation}\label{eq97}
{{\bar{h}}_{m}}\to {{h}_{m}}={{\bar{h}}_{m}}/\phi
\end{equation}
form a subgroup ${{W}^{g}}$ of the group $G{{L}^{g}}$ with parameters $L_{\ m}^{{{m}'}}=\phi \,\delta _{m}^{{{m}'}}$ \cite{Dirac}. So dilation gravity as a theory with Weyl symmetry ${{W}^{g}}$ is a special case of GRAF. In this case $\Delta \gamma _{\,\,\nu n}^{m}=-{{\partial }_{\nu }}\ln \phi \delta _{n}^{m}$, so $\Delta B_{m}^{\mu \nu }=0$, and formulas (\ref{eq63}) for the energy-momentum and suitable superpotential of gravitational field in DG give
\begin{equation}\label{eq98}
t_{\,m}^{\mu }=(\bar{t}_{m}^{\mu }+{{\partial }_{\sigma }}\ln \phi \,\bar{B}_{m}^{\mu \sigma })/\phi, \qquad
B_{\,m}^{\mu \sigma }=\bar{B}_{m}^{\mu \sigma }/\phi
\end{equation}
where $\bar{t}_{m}^{\mu }$ and $\bar{B}_{m}^{\mu \sigma }$ apply to the frame ${{\bar{h}}_{m}}$. In the case where ${{\bar{h}}_{m}}={{\partial }_{m}}$ is the coordinate frame of the Lagrangian coordinates ${{y}^{m}}$, the relation (\ref{eq98}) can be represented with the help of the Einstein energy-momentum complex $\bar{t}_{m}^{p}$ (\ref{eq91}) and Freud's superpotential $\bar{B}_{m}^{ps}$ (\ref{eq93}) by substituting in formulas (\ref{eq98}) $\bar{t}_{m}^{\mu }\to \bar{t}_{m}^{p}$, $\bar{B}_{m}^{\mu \sigma }\to \bar{B}_{m}^{ps}$ and entering the quantities $t_{m}^{\mu }\to t_{m}^{p}=\partial {{y}^{-1}}{{\partial }_{\mu }}{{y}^{p}}t_{m}^{\mu }$ and $B_{m}^{\mu \sigma }\to B_{m}^{ps}=\partial {{y}^{-1}}{{\partial }_{\mu }}{{y}^{p}}{{\partial }_{\sigma }}{{y}^{s}}B_{m}^{\mu \sigma }$: the energy-momentum complex and the corresponding superpotential of the gravitational field in DG in the Lagrangian coordinates ${{y}^{m}}$. In their terms, the gravitational field equation in DG takes the form ${{\partial }_{s}}B_{m}^{ps}=-t_{m}^{p}\,-\tau _{m}^{p}$, where $\tau _{m}^{p}=\bar{\tau }_{m}^{p}/\phi $.
	
The symmetry group of DG is $\tilde{S}_{M}^{g}=C{{}^{g}}\times\!\!\!|\, T_{M}^{g}$, where ${{C}^{g}}={{H}^{g}}\otimes {{W}^{g}}$.
	
The scalar field $\phi $ acts in DG as a finite parameter of dilation symmetry, so equation of motion with respect to $\phi $ is fulfilled automatically by the Palatini equation.

Moreover, Noether current and corresponding superpotential linked to Weyl transformations (\ref{eq97}) are zero due to the equality to zero of the current $J_{\{mn\}}^{\ \ \ \ \sigma }$  (\ref{eq57}).
	
Unimodular gravity is a theory where the determinant of the metric is fixed \cite{Unruh}. As mentioned above (in the subsection 3.3), in the GRAF the metric ${{g}_{mn}}$ in the affine frame ${{h}_{m}}$ can be considered as an arbitrary parameter of the theory (the metric of the background vacuum space), and therefore its determinant can be arbitrarily selected. In the case of a holonomic frame, that is, in EG, this limits the choice of Lagrangian coordinates by narrowing the EG symmetry subgroup ${{H}^{g}}$, associated with transitions between reference frames, to the group $SDiff$ of volume-preserving, or 'special' diffeomorphisms. It should be noted that the choice of Eulerian coordinates remains free, in particular, they can be dimensionless.

The symmetry group of UG is $\bar{S}_{\,M}^{vg}=SDiff\times\!\!\!|\, T_{M}^{g}$.

	The gravitational field equations in UG in Lagrangian coordinates can be represented in the same form (\ref{eq94}) as in EG, where we should fixed $v$ in expressions for the energy-momentum of the gravitational field $\bar{t}_{m}^{p}$ (\ref{eq91}) and superpotential $\bar{B}_{m}^{ps}$ (\ref{eq93}).

	Narrowing down $G{{L}^{g}}$-symmetry of GRAF to its other subgroups (classes of allowed affine frames), we can obtain other locally equivalent variants of GR.

\section{Discussion and Conclusions}

In this work, the method of deformations of generalized gauge groups is applied to the study of GR symmetry.
It is shown that all quantities which are conserved due to the invariance of the physical theories with respect to generalized gauge groups are quasilocal, that is, they have superpotentials.

GR is formulated in generalized reference frames, which are represented by (anholonomic in the general case) affine frame fields. The general principle of relativity is extended to the requirement of invariance of the theory with respect to transitions between generalized reference frames, that is, with respect to the group $G{{L}^{g}}$ of local linear transformations of affine frame fields.

GR is interpreted as the gauge theory of the gauge group of translations, and therefore is invariant under the space-time diffeomorphisms.

The consequence of the $G{{L}^{g}}$-invariance of the general relativity in an affine frame is the Palatini equation, which in the absence of torsion goes into the metricity condition, and vice versa, that is, is fulfilled identically in the Riemannian space.

The consequence of the $T_{M}^{g}$-invariance of GRAF is representation of the Einstein equation in superpotential form, that is, in the form of dynamic Maxwell equations (or Young-Mills equations).

At gauge translations, the energy-momentum of the gravitational field is transformed by the tensor law, and at transitions between relatively accelerated generalized reference frames - by the non-tensor law, which corresponds to the need to attract additional energy-momentum to provide such transitions.
The groups $G{{L}^{g}}$ and $T_{M}^{g}$ are united into group $S_{M}^{g}=G{{L}^{g}}\times\!\!\!|\, T_{M}^{g}$, which is their semidirect product and is the complete symmetry group of GRAF. Deformation of the group $S_{M}^{g}\to S_{M}^{gH}$ leads to renormalisation of energy-momentum of the gravitational field. Among the groups $S_{M}^{gH}$
there is one $DP$ - the group of parallel transports in the Riemannian space, action of which is agreed with the space geometrical structure.  The energy-momentum of the gravitational field for $DP$ is localizable.

Limiting the admissible generalized reference frames by imposing $G{{L}^{g}}$-gauge conditions on field variables of GRAF leads to various locally equivalent formulations of GR, such as EG, GROF or TEGR, DG, UG, and others. However, these theories can significantly be different by global solutions. So in \cite{Gielen} was suggested that "just like a conformally invariant theory of matter is insensitive to the cosmological singularity, a $G{{L}^{g}}$-invariant field theory will be insensitive to any singularity". This hope is based on a sufficiently broad gauge symmetry of such theory, which can provide the possibility of finding generalized reference frames in which singularities will be absent, just as singularity on the event horizon in the Schwarzschild solution  disappears in the free-falling holonomic reference frame.

The choice of appropriate generalized d-dimensional reference frames in the framework of group-theory approach to unification of gravity with internal symmetry gauge interactions \cite{Sam5}, \cite{Sam6} gives hope for getting rid of singularities in other theories of gauge interactions, too.

We are planning to consider the physical arguments regarding the choice of generalized reference frames in the future.

\section*{Acknowledgements}
The author is grateful to Alisa Gryshchenko for her help with the prepare of article.

 Chair of Applied Mathematics \\
 Dniprovsk State Technical University, UKRAINE \\
{\it E-mail address}: {\tt serh.samokhval@gmail.com}

\label{last}
\end{document}

%% file: somedef.tex
\theoremstyle{plain}